\documentclass[granma,referee]{svjour}
\usepackage{amsmath}
\usepackage{graphics}
\begin{document}
\title{Computer simulation of uniformly heated granular fluids}
\author{Jos\'e Mar\'{\i}a Montanero, Andr\'es Santos
}                     
\offprints{J. M. Montanero}          
\mail{A. Santos} \institute{Departamento de Electr\'onica e
Ingenier\'{\i}a Electromec\'anica, Universidad de Extremadura,
E-06071 Badajoz, Spain\\ \email{jmm@unex.es} \\ Departamento de
F\'{\i}sica Universidad de Extremadura, E-06071 Badajoz, Spain \\
\email{andres@unex.es}}
\date{Received: 17 June 1999}
%
\maketitle
\begin{abstract}
Direct Monte Carlo simulations of the Enskog-Boltzmann equation for a
spatially uniform system of smooth inelastic spheres are performed.
In order to reach a steady state, the particles are assumed to be
under  the action of an external driving force which does work  to
compensate for the collisional loss of energy. Three different types of
external driving are
considered: (a) a stochastic force, (b) a deterministic force
proportional to the particle velocity and (c) a deterministic force
parallel to the particle velocity but constant in magnitude. The
Enskog-Boltzmann equation in case (b) is fully equivalent to that of the
homogeneous cooling state (where the thermal velocity monotonically
decreases with time) when expressed in
terms of the particle velocity relative to the thermal velocity. Comparison
of the simulation results for the fourth cumulant and the high energy tail
with theoretical predictions derived in cases (a) and (b)  [T. P. C. van
Noije and M. H. Ernst, Gran. Matt. 1, 57 (1998)] shows a good agreement.
In contrast to these two cases,
  the deviation from the
Maxwell-Boltzmann distribution is not well represented by Sonine
polynomials in case (c), even for low dissipation. In addition, the high energy tail
exhibits an underpopulation effect in this case.
\end{abstract}
\section{Introduction}
\label{sec:1}
Most of the recent studies of rapid granular flow \cite{C90} are based on
the Enskog equation for the velocity distribution function $f(\mathbf{r},
\mathbf{v},t)$ of an assembly of inelastic hard spheres \cite{BDS97}.
In the special case of a spatially uniform state, the Enskog equation reads
\begin{equation}
\label{1}
\frac{\partial}{\partial t} f(\mathbf{v}_1,t)+\mathcal{F}f(\mathbf{v}_1,t)=
\chi I[\mathbf{v}_1|f(t),f(t)],
\end{equation}
where
\begin{eqnarray}
\label{2}
I[\mathbf{v}_1|f(t),f(t)]&\equiv&\sigma^{d-1}\int d\mathbf{v}_2\int
d\widehat{\boldsymbol{\sigma}}\,
\Theta(\mathbf{v}_{12}\cdot\widehat{\boldsymbol{\sigma}})
(\mathbf{v}_{12}\cdot\widehat{\boldsymbol{\sigma}})\nonumber\\
&&\times\left[\alpha^{-2}f(\mathbf{v}_1'',t)f(\mathbf{v}_2'',t)-
f(\mathbf{v}_1,t)f(\mathbf{v}_2,t)\right].\nonumber\\
&&
\end{eqnarray}
In Eq.\ (\ref{1}) $\mathcal{F}$ is an operator representing the effect of
an external force (if it exists), $\chi$ is the pair correlation function at
contact and $I$ is the collision operator. In  Eq.\
(\ref{2}), $d$ is the dimensionality of the system, $\sigma$ is the diameter
of the spheres, $\mathbf{v}_{12}\equiv \mathbf{v}_1-\mathbf{v}_2$ is the
relative velocity of the colliding particles,
$\widehat{\boldsymbol{\sigma}}$ is a unit vector directed along the line of
centers from the sphere $1$ to the sphere $2$, $\Theta$ is the Heaviside
step function and $\alpha<1$ is the coefficient of normal restitution, here
assumed to be constant. In addition, $(\mathbf{v}_1'',\mathbf{v}_2'')$ are
the precollisional velocities yielding $(\mathbf{v}_1,\mathbf{v}_2)$ as the
postcollisional ones, i.e.
$\mathbf{v}_{1,2}''=\mathbf{v}_{1,2}\mp\frac{1}{2}(1+\alpha^{-1})
(\mathbf{v}_{12}\cdot\widehat{\boldsymbol{\sigma}})
\widehat{\boldsymbol{\sigma}}$.
Except  for the presence of the factor $\chi$, which accounts for the
increase of the collision frequency due to excluded volume effects, the
Enskog equation for uniform states, Eq.\ (\ref{1}), becomes identical
with the Boltzmann equation.

In the case of elastic particles ($\alpha=1$) and in the absence of external
forcing ($\mathcal{F}=0$), it is well known that the long-time solution of
Eq.\ (\ref{1}) is the Maxwell-Boltzmann equilibrium distribution function,
$f(\mathbf{v},t)\to n v_0^{-d}\phi(\mathbf{v}/v_0)$,
$\phi(\mathbf{c})\equiv\pi^{-d/2}e^{-c^2}$, where $n$ is the number density,
$v_0=(2\langle v^2\rangle/d)^{1/2}$ is the thermal velocity and
$\mathbf{c}=\mathbf{v}/v_0$ is the reduced velocity. On the other hand, if
the particles are inelastic ($\alpha<1$) and $\mathcal{F}=0$, a steady
state is not possible in uniform situations since, due to the
dissipation of energy through collisions, the thermal velocity $v_0(t)$
decreases monotonically with time. Regardless of the initial uniform
state, the solution of Eq.\ (\ref{1}) tends to the so-called homogeneous
cooling state \cite{GS95,NE96,NE98,BMC96}, characterized by the fact that
the time dependence occurs only through the thermal velocity $v_0(t)$:
$f(\mathbf{v},t)\to n v_0^{-d}(t)\widetilde{f}(\mathbf{v}/v_0(t))$. In
addition, $\widetilde{f}(\mathbf{c})$ deviates from a Maxwellian,
$\widetilde{f}(\mathbf{c})\neq\phi(\mathbf{c})$, as measured by the fourth
cumulant
\begin{equation}
\label{3}
a_2\equiv\frac{d}{d+2}\frac{\langle v^4\rangle}{\langle v^2\rangle^2}-1
=\frac{4}{d(d+2)}\langle c^4\rangle-1,
\end{equation}
where
\begin{equation}
\label{4}
\langle c^p\rangle\equiv \int d\mathbf{c}\,c^p\widetilde{f}(\mathbf{c}).
\end{equation}
By expanding $\widetilde{f}(\mathbf{c})/\phi(\mathbf{c})$ in a set of Sonine
polynomials $\{S_p(c^2)\}$ and neglecting the terms beyond $p=2$, van Noije
and Ernst \cite{NE96,NE98} have estimated the value of $a_2$:
\begin{equation}
\label{5}
a_2(\alpha)\simeq\frac{16(1-\alpha)(1-2\alpha^2)}{9+24d-\alpha(41-8
d)+30(1-\alpha)\alpha^2}.
\end{equation}
The above expression corrects an algebraic error in a previous calculation
of $a_2$ in the three-dimensional case \cite{GS95}.
According to Eq.\ (\ref{5}), $a_2$ changes sign at $\alpha=1/\sqrt{2}\simeq
0.71$.
By using the same
method, Garz\'o and Dufty \cite{GD99} have recently extended the
evaluation of $a_2$ to a binary mixture of hard spheres.
The accuracy of Eq.\ (\ref{5}) has been quantitatively confirmed by Brey et
al. \cite{BMC96} from Monte Carlo simulations of the Boltzmann equation for
hard spheres ($d=3$) in the range $0.7\leq \alpha\leq 1$.
As a complementary measure of the departure of $\widetilde{f}(\mathbf{c})$
from $\phi(\mathbf{c})$, Esipov and P\"oschel \cite{EP97} and van Noije and
Ernst \cite{NE98} have analyzed the high energy tail of the distribution
function and have found an asymptotic behavior of the form
\begin{equation}
\label{5bis}
\log \widetilde{f}(\mathbf{c})\sim -c,
\end{equation}
 in contrast to
$\log \phi(\mathbf{c})\sim -c^2$.
The high energy tail (\ref{5bis}) has been confirmed by simulations in the
case of hard disks ($d=2$) \cite{BCR99}.

In order to reach a  steady state,
energy injection  is needed to compensate for the energy dissipated
through collisions.  This can be achieved by vibration
 of vessels \cite{ER89} or in fluidized beds \cite{IH95}.
The same effect can be obtained by means of  external driving forces acting
locally on each particle \cite{WM96}.
Borrowing a terminology frequently used in nonequilibrium molecular
dynamics of  elastic particles \cite{EM90}, we will call
 this type of external forces ``thermostats''.
In general, the equation of motion for a particle $i$ is then
\begin{equation}
\label{6}
m\dot{\mathbf{v}}_i=
\mathbf{F}_i^{\mbox{\scriptsize coll}}+
\mathbf{F}_i^{\mbox{\scriptsize th}},
\end{equation}
where $m$ is the mass of a particle, $\mathbf{F}_i^{\mbox{\scriptsize
coll}}$ is the force due to collisions and $\mathbf{F}_i^{\mbox{\scriptsize
th}}$ is the thermostat force. Williams and MacKintosh \cite{WM96}
introduced a stochastic force assumed to have the form of a Gaussian white
noise:
\begin{equation}
\label{7}
\langle
\mathbf{F}_i^{\mbox{\scriptsize th}}(t)\rangle =\mathbf{0},\quad
\langle
\mathbf{F}_i^{\mbox{\scriptsize th}}(t)
\mathbf{F}_j^{\mbox{\scriptsize th}}(t')\rangle
=\mathsf{I}m^2\xi_0^2  \delta_{ij}\delta(t-t'),
\end{equation}
where $\mathsf{I}$ is the $d\times d$ unit matrix and $\xi_0^2$ represents
the strength of the correlation. The corresponding operator $\mathcal{F}$
appearing in Eq.\ (\ref{1}) has a Fokker-Planck form \cite{NE98}:
\begin{equation}
\label{8}
\mathcal{F}f(\mathbf{v}_1)=-\frac{\xi_0^2}{2}\left(\frac{\partial}{\partial
\mathbf{v}_1}\right)^2f(\mathbf{v}_1).
\end{equation}
Van Noije and Ernst \cite{NE98} have studied the stationary solution of the
uniform equation (\ref{1}) with the thermostat (\ref{8}). They have
found for the coefficient $a_2$ defined by Eq.\ (\ref{3}) the value
\begin{equation}
\label{9}
a_2(\alpha)\simeq\frac{16(1-\alpha)(1-2\alpha^2)}{73+56d-3\alpha (35+8
d)+30(1-\alpha)\alpha^2}.
\end{equation}
The high energy tail is \cite{NE98}
\begin{equation}
\label{10}
\log \widetilde{f}(\mathbf{c})\sim -c^{3/2}.
\end{equation}

Of course, deterministic thermostats can also be used. For instance, the use
of Gauss's principle of least constraint leads to the thermostat force
\cite{EM90}
\begin{equation}
\label{11}
\mathbf{F}_i^{\mbox{\scriptsize th}}=m\zeta \mathbf{v}_i,
\end{equation}
where $\zeta$ is a positive constant. In this case,
\begin{equation}
\label{12}
\mathcal{F}f(\mathbf{v}_1)=\zeta\frac{\partial}{\partial
\mathbf{v}_1}\cdot\left[\mathbf{v}_1f(\mathbf{v}_1)\right].
\end{equation}
It is interesting pointing out that the Enskog-Boltzmann equation (\ref{1})
for the above Gaussian thermostat force is formally identical with the
equation for the homogeneous cooling state (i.e. with $\mathcal{F}=0$) when
both equations are expressed in terms of the reduced distribution
$\widetilde{f}(\mathbf{c})$ (see Sect.\ \ref{sec:2}). As a consequence, the
results (\ref{5}) and (\ref{5bis}) apply to this thermostatted case as well.

The differences between Eqs.\ (\ref{5}) and (\ref{9}) and between
(\ref{5bis}) and (\ref{10}) illustrate the influence of the thermostat force
on the departure of the steady-state distribution function from the
Maxwell-Boltzmann distribution. In the case of the stochastic force, Eq.\
(\ref{8}), $\widetilde{f}(\mathbf{c})$ is closer to $\phi(\mathbf{c})$ than
in the case of the Gaussian force, Eq.\ (\ref{12}), since
in the former case  $a_2$ and the high energy overpopulation are
smaller than in the latter case.
Of course, other types of thermostats are also possible. For example,
a different choice for a deterministic
thermostat is
\begin{equation}
\label{13}
\mathbf{F}_i^{\mbox{\scriptsize th}}=mg \widehat{\mathbf{v}}_i,
\end{equation}
where $\widehat{\mathbf{v}}_i\equiv{\mathbf{v}}_i/v_i$.
While the Gaussian force, Eq.\ (\ref{11}), is proportional to the velocity
of the particle, Eq.\ (\ref{13}) corresponds to a force that is parallel to
the direction of motion but constant in magnitude. The corresponding
operator $\mathcal{F}$ is
\begin{equation}
\label{14}
\mathcal{F}f(\mathbf{v}_1)=\frac{g}{v_1}\left\{\frac{\partial}{\partial
\mathbf{v}_1}\cdot\left[\mathbf{v}_1f(\mathbf{v}_1)\right]-
f(\mathbf{v}_1)\right\}.
\end{equation}

The aim of this paper is to present direct  Monte Carlo simulations of Eq.\
(\ref{1}) with the three choices for the thermostat, Eqs.\ (\ref{8}),
(\ref{12}) and (\ref{14}). In the cases of the stochastic and the
Gaussian thermostats, we will confirm the tails (\ref{10}) and (\ref{5bis})
and will check the accuracy of the estimates (\ref{9}) and (\ref{5}). In the
latter case, however, we will see that a better agreement with simulation
results for $\alpha< 0.5$ is obtained if an estimate slightly different from
(\ref{5}) is used. The simulation results corresponding to the non-Gaussian
thermostat (\ref{14}) show that, in contrast to what happens in the two
previous cases, $a_2$ remains negative for all $\alpha$. This
feature is qualitatively captured by an estimate derived from a
Sonine approximation. In this problem, however, the Sonine polynomials do
not constitute a good set for the expansion of
$\widetilde{f}(\mathbf{c})/\phi(\mathbf{c})$ and, consequently, the estimate
is not quantitatively good. Besides, the high energy tail is of the form
$\log \widetilde{f}(\mathbf{c})\sim -c^2$, but with a coefficient different
from that of the Maxwell-Boltzmann distribution.

The organization of this paper is as follows. The theoretical analysis is
reviewed in Sect.\ \ref{sec:2}. The computer simulation method employed
to solve numerically the uniform Enskog-Boltzmann equation is described in
Sect.\ \ref{sec:3}. The results are presented and compared with the
theoretical predictions in Sect.\ \ref{sec:4}.
The paper ends with a
summary and discussion in Sect.\ \ref{sec:5}.

\section{Theoretical predictions}
\label{sec:2}

In the steady state, the Enskog-Boltzmann equation (\ref{1}) can be
expressed in terms of the reduced velocity distribution function
$\widetilde{f}(\mathbf{c})$ as
\begin{equation}
\label{15}
\widetilde{\mathcal{F}}\widetilde{f}(\mathbf{c}_1)=
\chi \widetilde{I}[\mathbf{c}_1|\widetilde{f},\widetilde{f}],
\end{equation}
where
\begin{eqnarray}
\label{16}
\widetilde{I}[\mathbf{c}_1|\widetilde{f},\widetilde{f}]&\equiv&\int
d\mathbf{c}_2\int d\widehat{\boldsymbol{\sigma}}\,
\Theta(\mathbf{c}_{12}\cdot\widehat{\boldsymbol{\sigma}})
(\mathbf{c}_{12}\cdot\widehat{\boldsymbol{\sigma}})\nonumber\\
&&\times\left[\alpha^{-2}\widetilde{f}(\mathbf{c}_1'')
\widetilde{f}(\mathbf{c}_2'')-
\widetilde{f}(\mathbf{c}_1)\widetilde{f}(\mathbf{c}_2)\right].
\end{eqnarray}
The reduced operator $\widetilde{\mathcal{F}}$ for the stochastic [Eq.\
(\ref{8})], Gaussian [Eq.\ (\ref{12})] and non-Gaussian [Eq.\ (\ref{14})]
thermostats,  is
\begin{equation}
\label{17}
\widetilde{\mathcal{F}}\widetilde{f}(\mathbf{c}_1)=
-\frac{\xi_0^2}{2v_0^3
n\sigma^{d-1}}c_1^{-(d-1)}
\frac{\partial}{\partial
{c}_1}\left[c_1^{d-1}\frac{\partial \widetilde{f}(\mathbf{c}_1)}{\partial
c_1}\right],
\end{equation}
\begin{equation}
\label{18}
\widetilde{\mathcal{F}}\widetilde{f}(\mathbf{c}_1)=
\frac{\zeta}{v_0n\sigma^{d-1}}
c_1^{-(d-1)}
\frac{\partial}{\partial
{c}_1}\left[c_1^{d}\widetilde{f}(\mathbf{c}_1)\right],
\end{equation}
\begin{equation}
\label{19}
\widetilde{\mathcal{F}}\widetilde{f}(\mathbf{c}_1)=
\frac{g}{v_0^2n\sigma^{d-1}}
c_1^{-(d-1)}
\frac{\partial}{\partial
{c}_1}\left[c_1^{d-1}\widetilde{f}(\mathbf{c}_1)\right],
\end{equation}
respectively. In Eqs.\ (\ref{17})--(\ref{19}) we have already taken into
account that the distribution function must be isotropic in the steady state.
Equation (\ref{15}) with the term (\ref{18}) is fully equivalent to Eq.\
(10) of Ref.\ \cite{NE98}, the latter being derived in the context of the
homogeneous cooling state. This formal equivalence between the free evolving
state and the one controlled by a Gaussian external force is also present in
the case of elastic particles interacting via arbitrary power-law
potentials in homogeneous situations \cite{GSB90} or via the Maxwell
potential in  the uniform shear flow \cite{DSBR86}.

\subsection{Stochastic thermostat}
\label{sec:2.1}
For the sake of completeness,  we summarize now some of the results obtained
in Ref.\ \cite{NE98}.
In order to characterize the deviation of $\widetilde{f}(\mathbf{c})$ from
$\phi(\mathbf{c})$ by means of the cumulant (\ref{3}), it is useful to
consider the hierarchy of moment equations. Multiplying both sides of Eq.\
(\ref{15}) by $c_1^p$ and integrating over $\mathbf{c}_1$, we get
\begin{equation}
\label{20}
{\mu_p}={\mu_2}\frac{p(p+d-2)}{2d}\langle c^{p-2}\rangle
\end{equation}
for the stochastic thermostat,
where we have defined
\begin{equation}
\label{21} \mu_p\equiv - \int
d\mathbf{c}\,c^p\widetilde{I}[\mathbf{c}|\widetilde{f},\widetilde{f}].
\end{equation}
In Eq.\ (\ref{20}) we
have taken into account the normalization condition $\langle c^{0}\rangle
=1$, so that
$\mu_2=d{\xi_0^2}/{v_0^3\chi n\sigma^{d-1}}$.
In the special case of $p=4$, Eq.\ (\ref{20}) becomes
\begin{equation}
\label{22}
{\mu_4}=(d+2){\mu_2},
\end{equation}
where we have used the fact that, by definition,
$\langle c^{2}\rangle=d/2$.
Equations (\ref{21}) and (\ref{22}) are still exact. To get an approximate
expression for $a_2(\alpha)$, three steps will be taken \cite{NE98}. First,
we assume that $\widetilde{f}$ can be well described by the simplest
Sonine approximation, at least for the velocities relevant to the evaluation
of $a_2$. Thus,
\begin{equation}
\label{23}
\widetilde{f}(\mathbf{c})\simeq\phi(\mathbf{c})\left[1+a_2 S_2(c^2)\right],
\end{equation}
where
\begin{equation}
\label{24}
S_2(x)=\frac{1}{2}x^2-\frac{d+2}{2}x+\frac{d(d+2)}{8}.
\end{equation}
The approximation (\ref{23}) is justified by the fact that $a_2$ is expected
to be small. The second step consists of inserting Eq.\ (\ref{23}) into Eq.\
(\ref{21}) and neglecting terms nonlinear in $a_2$. For $\mu_2$ and $\mu_4$
the results are \cite{NE98}
\begin{equation}
\label{25}
\mu_p\simeq\mu_p^{(0)}+\mu_p^{(1)}a_2,
\end{equation}
with
\begin{equation}
\label{26}
\mu_2^{(0)}\equiv\frac{\pi^{(d-1)/2}}{\sqrt{2}\Gamma(d/2)}(1-\alpha^2),
\end{equation}
\begin{equation}
\label{27}
\mu_2^{(1)}\equiv\frac{3}{16}\mu_2^{(0)},
\end{equation}
\begin{equation}
\label{28}
\mu_4^{(0)}\equiv\left(d+\frac{3}{2}+\alpha^2\right)\mu_2^{(0)},
\end{equation}
\begin{equation}
\label{29}
\mu_4^{(1)}\equiv\left[\frac{3}{32}\left(10d+39+10\alpha^2\right)
+\frac{d-1}{1-\alpha}\right]\mu_2^{(0)}.
\end{equation}
In the third step, the approximations (\ref{25}) with $p=2 \mbox{ and }4$
are inserted into the exact equation (\ref{22})  and $a_2$ is
obtained from the resulting linear equation:
\begin{equation}
\label{30}
a_2\simeq -\frac{\mu_4^{(0)}-(d+2)\mu_2^{(0)}}{\mu_4^{(1)}-(d+2)\mu_2^{(1)}}.
\end{equation}
This is the result derived by van Noije and Ernst \cite{NE98}, Eq.\
(\ref{9}).
It must be pointed  out that a certain
degree of ambiguity is present in this last step. For instance, if
Eq.\ (\ref{22}) were written as $\mu_4/\mu_2=d+2$, we could expand the ratio
$\mu_4/\mu_2$ in powers of $a_2$ and neglect nonlinear terms to find
\begin{eqnarray}
\label{31}
a_2&\simeq& -
\frac{\mu_4^{(0)}-(d+2)\mu_2^{(0)}}{\mu_4^{(1)}-
\mu_2^{(1)}\mu_4^{(0)}/\mu_2^{(0)}}\nonumber\\
&=&
\frac{4(1-\alpha)(1-2\alpha^2)}{19+14d-3\alpha (9+2
d)+6(1-\alpha)\alpha^2}.
\end{eqnarray}
However, since $a_2$ is indeed small ($|a_2|<0.1$), Eqs.\ (\ref{9}) and
(\ref{31}) give practically identical results, the maximum deviation being
less than about 0.001.

Now we consider the high energy tail.
In general, the collision integral can be decomposed into a gain and a loss
term: $\widetilde{I}[\mathbf{c}_1|\widetilde{f},\widetilde{f}]=
\widetilde{I}_g[\mathbf{c}_1|\widetilde{f},\widetilde{f}]-
\widetilde{I}_l[\mathbf{c}_1|\widetilde{f},\widetilde{f}]$.
For large $c_1$ the loss term can be approximated as
\begin{eqnarray}
\label{32}
\widetilde{I}_l[\mathbf{c}_1|\widetilde{f},\widetilde{f}]
&=&\beta_1
\int
d\mathbf{c}_2\, c_{12}
\widetilde{f}(\mathbf{c}_1)\widetilde{f}(\mathbf{c}_2)\nonumber\\
&\approx&
\beta_1 c_1 \widetilde{f}(\mathbf{c}_1),
\end{eqnarray}
where $\beta_1$ is defined by Eq.\ (\ref{A5}).
Let us assume that for large velocities the gain term is negligible
versus the loss term, i.e.
\begin{equation}
\label{32bis}
\lim_{c_1\to \infty}
\frac{\widetilde{I}_g[\mathbf{c}_1|\widetilde{f},\widetilde{f}]}{
\widetilde{I}_l[\mathbf{c}_1|\widetilde{f},\widetilde{f}]}=0.
\end{equation}
In that case,
the Enskog-Boltzmann equation for the stochastic
thermostat becomes
\begin{equation}
\label{33}
\frac{\mu_2}{2d}c^{-(d-1)}
\frac{\partial}{\partial
{c}}\left[c^{d-1}\frac{\partial \widetilde{f}(\mathbf{c})}{\partial
c}\right]\approx
\beta_1 c \widetilde{f}(\mathbf{c}).
\end{equation}
The solution of this equation for large $c$ is
\begin{equation}
\label{34}
\widetilde{f}(\mathbf{c})\approx K \exp\left(-Ac^{3/2}\right),\quad
A\equiv\frac{2}{3}\left(\frac{2d\beta_1}{\mu_2}\right)^{1/2},
\end{equation}
where $K$ is an undetermined constant.
By arguments given in Ref.\
\cite{NE98}, it can be seen that the result (\ref{34}) is indeed consistent
with the assumption (\ref{32bis}).
Equation (\ref{34}) shows an \textit{overpopulation\/} with respect to
the Maxwell-Boltzmann tail. On the other hand, as $\alpha\to 1$, the
amplitude $A$ diverges as $(1-\alpha)^{-1/2}$, thus indicating that the
overpopulation effect is restricted to  larger and larger
energies in the limit $\alpha\to 1$.

\subsection{Gaussian thermostat}
\label{sec:2.2}
In the case of the deterministic Gaussian thermostat, Eq.\ (\ref{18}), the
moment equation is
\begin{equation}
\label{35}
\mu_p=\mu_2 \frac{p}{d}\langle c^p\rangle,
\end{equation}
where now
$\mu_2=d{\zeta}/{v_0\chi n\sigma^{d-1}}$.
 If we set $p=4$,
\begin{equation}
\label{37}
\mu_4=(d+2)(1+a_2)\mu_2,
\end{equation}
where we have made use of Eq.\ (\ref{3}).
Substituting the approximation (\ref{25}) and neglecting terms nonlinear in
$a_2$, we get
\begin{equation}
\label{38}
a_2\simeq -\frac{\mu_4^{(0)}-(d+2)\mu_2^{(0)}}{\mu_4^{(1)}-(d+2)
(\mu_2^{(1)}+\mu_2^{(0)})},
\end{equation}
which is the same as Eq.\ (\ref{5}).
There exists again some arbitrariness about the use of the exact
equation (\ref{37}) in connection with the approximation (\ref{25}). If we
rewrite (\ref{37}) as $\mu_4/\mu_2=(d+2)(1+a_2)$ and neglect nonlinear
terms, the resulting $a_2$ is fairly close to Eq.\ (\ref{38}). On the other
hand, if we start from $\mu_4/(1+a_2)=(d+2)\mu_2$, the result is
\begin{eqnarray}
\label{39}
a_2&\simeq& -\frac{\mu_4^{(0)}-(d+2)\mu_2^{(0)}}{\mu_4^{(1)}-
\mu_4^{(0)}-(d+2)
\mu_2^{(1)}}\nonumber\\
&=&
\frac{16(1-\alpha)(1-2\alpha^2)}{25+24d-\alpha (57-
8d)-2(1-\alpha)\alpha^2}.
\end{eqnarray}
The estimates (\ref{5}) and (\ref{39}) practically coincide in the region
$0.5<\alpha\leq 1$. However, they visibly separate for larger dissipation.
In the interval $0<\alpha<0.3$, the values given by
Eq.\ (\ref{5}) are 12\%--20\% ($d=3$) or 18\%--28\% ($d=2$) larger than
those given by Eq.\ (\ref{39}). As we will see later, the simulation results
indicate that Eq.\ (\ref{39}) is a better estimate than Eq.\ (\ref{5}).

For large $c$ the Enskog-Boltzmann equation becomes
\begin{equation}
\label{40}
\frac{\mu_2}{d}c^{-(d-1)}
\frac{\partial}{\partial
{c}}\left[c^{d}\widetilde{f}(\mathbf{c})\right]
\approx
-\beta_1 c \widetilde{f}(\mathbf{c}),
\end{equation}
where we have used Eqs.\ (\ref{32}) and (\ref{32bis}).
Its solution is
\begin{equation}
\label{41}
\widetilde{f}(\mathbf{c})\approx K \exp\left(-Ac\right),\quad
A\equiv\frac{d\beta_1}{\mu_2}.
\end{equation}
Again, this result is seen to be consistent with (\ref{32bis}) \cite{NE98}.
Equation (\ref{41}) indicates an overpopulation effect even larger than
with the stochastic thermostat.

\subsection{Non-Gaussian thermostat}
\label{sec:2.3}
Now we consider the deterministic non-Gaussian thermostat (\ref{13}),
represented by the operator (\ref{19}). To the best of our knowledge, this
external force has not been analyzed before. The corresponding moment
equation is
\begin{equation}
\label{42}
\mu_p=\mu_2 \frac{p}{2}\frac{\langle c^{p-1}\rangle}{\langle c\rangle},
\end{equation}
where
$\mu_2={2g}{\langle c\rangle}/{v_0^2\chi n\sigma^{d-1}}$.
In particular,
\begin{equation}
\label{43bis}
\mu_4=2\mu_2 \frac{\langle c^{3}\rangle}{\langle c\rangle}.
\end{equation}
In contrast to the two previous cases, now the even  collisional
moments $\mu_p$ are coupled to the odd moments $\langle
c^{p-1}\rangle$, and vice versa. In terms of the energy variable
$\epsilon=c^2$, this means that the integer collisional moments are coupled
to the half-integers energy moments. This is  related to the fact that the
force (\ref{13}) is singular at $\epsilon=0$.
As a consequence, while $\widetilde{f}(\mathbf{c})$ is expected to be close
to the Maxwellian $\phi(\mathbf{c})$, the ratio
$\widetilde{f}(\mathbf{c})/\phi(\mathbf{c})$ is singular at
$\epsilon=0$ and thus it is not well represented by an expansion in
$\{S_p(\epsilon)\}$.
To be more precise, let us define the function $\Delta(c)$ by the equation
\begin{equation}
\label{44}
\widetilde{f}(\mathbf{c})=\phi(\mathbf{c})\left[1+a_2 \Delta(c)\right].
\end{equation}
Therefore,
\begin{equation}
\label{45}
\int d\mathbf{c}\,\phi(\mathbf{c})\Delta(c)=
\int d\mathbf{c}\,\phi(\mathbf{c})c^2\Delta(c)=0,
\end{equation}
\begin{equation}
\label{46}
\int d\mathbf{c}\,\phi(\mathbf{c})c^4\Delta(c)=\frac{d(d+2)}{4}.
\end{equation}
The polynomial $S_2(c^2)$ verifies the above equalities.
As a matter of fact,  $\Delta(c)\simeq S_2(c^2)$ in the cases of the
thermostats (\ref{17}) and (\ref{18}). This is not so, however, in the case
of (\ref{19}), even in the limit of low dissipation. As we will see in
Sect.\ \ref{sec:4}, $\left. \partial \Delta(c)/\partial c\right|_{c=0}\neq
0$, what indicates that $\Delta(c)$ is essentially different from a
polynomial in $c^2$.
All of this complicates the evaluation of $a_2$. Nevertheless, since
$\Delta(c)$ and $S_2(c^2)$ share the moments of degrees 0, 2 and 4 [cf.
Eqs.\ (\ref{45}) and (\ref{46})], we can expect to obtain a {\em crude\/}
estimate of $a_2$ by assuming that in the calculation of $\langle c\rangle$,
 $\langle c^3\rangle$, $\mu_2$ and $\mu_4$ we can replace $\Delta(c)$ by
$S_2(c^2)$. If  that were the case, $\mu_2$ and $\mu_4$ would be given by
Eqs.\ (\ref{25})--(\ref{29}) and
\begin{equation}
\label{47}
\langle c\rangle\simeq \frac{\Gamma((d+1)/2)}{\Gamma(d/2)}
\left(1-\frac{1}{8}a_2\right),
\end{equation}
\begin{equation}
\label{47bis}
\langle c^3\rangle\simeq \frac{\Gamma((d+3)/2)}{\Gamma(d/2)}
\left(1+\frac{3}{8}a_2\right).
\end{equation}
Inserting this into Eq.\ (\ref{43bis})  and neglecting
nonlinear terms, we get
\begin{eqnarray}
\label{48}
a_2&\simeq& -\frac{\mu_4^{(0)}-(d+1)\mu_2^{(0)}}{\mu_4^{(1)}-
(d+1)(
\mu_2^{(1)}+\mu_2^{(0)}/2)}\nonumber\\
&=&-
\frac{16(1-\alpha)(1+2\alpha^2)}{63+40d-\alpha (95+
8d)+30(1-\alpha)\alpha^2}.
\end{eqnarray}
While in the cases of the stochastic thermostat, Eqs.\ (\ref{9}) or
(\ref{31}), and the Gaussian thermostat, Eqs.\ (\ref{5}) or (\ref{39}), the
cumulant $a_2$ changes from negative to positive values at $\alpha\simeq
0.71$, Eq.\ (\ref{48}) indicates that $a_2$ remains negative in the case of
the non-Gaussian thermostat. We will see in Sect.\ \ref{sec:4} that our
computer simulations confirm this feature. At a quantitative level, however,
the estimate (\ref{48}) is about 20\% too small in magnitude.

To analyze the high energy tail, let us assume for the moment the validity
of (\ref{32bis}), so that the Enskog-Boltzmann equation can be
replaced by
\begin{equation}
\label{49}
\frac{\mu_2}{2\langle c\rangle}c^{-(d-1)}
\frac{\partial}{\partial
{c}}\left[c^{d-1}\widetilde{f}(\mathbf{c})\right]
\approx
-\beta_1 c \widetilde{f}(\mathbf{c}),
\end{equation}
whose solution for large $c$ is
\begin{equation}
\label{50}
\widetilde{f}(\mathbf{c})\approx K \exp\left(-A'c^2\right),\quad
A'\equiv\frac{\beta_1\langle c\rangle}{\mu_2}.
\end{equation}
According to (\ref{50}), $\widetilde{f}(\mathbf{c})$ has a Maxwellian tail
that is \textit{underpopulated\/} with respect to the Maxwell-Boltzmann
distribution $\phi(\mathbf{c})$, since the amplitude $A'\simeq
\sqrt{2}/(1-\alpha^2)$ is larger than 1. But now we get an unphysical
result: the underpopulation effect increases as one approaches
the elastic limit, since $A'\to\infty$ as $\alpha\to 1$. The solution to
this paradox lies in the fact that the assumption (\ref{32bis}) is not
justified in this case. Let us assume instead that the gain and loss term
are comparable, namely
\begin{equation}
\label{51}
\widetilde{I}_g[\mathbf{c}_1|\widetilde{f},\widetilde{f}]
\approx
\gamma\beta_1 c_1 \widetilde{f}(\mathbf{c}_1),
\end{equation}
where $\gamma<1$ is an unknown function of $\alpha$.
According to this, Eq.\ (\ref{50}) is replaced by
\begin{equation}
\label{52}
\widetilde{f}(\mathbf{c})\approx K \exp\left(-Ac^2\right),\quad
A\equiv A'(1-\gamma).
\end{equation}
On physical grounds we expect that $A\to 1$ when $\alpha\to 1$, which
implies that $\gamma\to 1-\sqrt{2}(1-\alpha)$ in that limit.
As will be shown in Sect.\ \ref{sec:4}, comparison with simulation results
confirms a behavior of the form (\ref{52}).

\section{Direct Simulation Monte Carlo method}
\label{sec:3}
The Direct Simulation Monte Carlo  (DSMC) method devised by Bird \cite{Bird}
has proven to be a very efficient tool to solve numerically the Boltzmann
equation. The DSMC method has been recently extended to the Enskog equation
\cite{MS96} and its application to inelastic particles is straightforward
\cite{BMC96,MGSB99}.
Here we briefly describe the specific method we have used to solve the
uniform Enskog-Boltzmann equation (\ref{1}) in the case of a
three-dimensional system ($d=3$).

The velocity distribution function is represented by the velocities
$\{\mathbf{v}_i\}$ of $N$ ``simulated'' particles:
\begin{equation}
\label{53}
f(\mathbf{v},t)\to n\frac{1}{N}\sum_{i=1}^N
\delta\left(\mathbf{v}_i(t)-\mathbf{v}\right).
\end{equation}
At the initial state the particles are assigned velocities drawn from a
Maxwell-Boltzmann probability distribution:
\begin{equation}
\label{53bis}
n^{-1}f(\mathbf{v},0)=\pi^{-3/2} v_{0}^{-3}(0) e^{-v^2/v_{0}^2(0)},
\end{equation}
where $v_{0}(0)$ is an arbitrary initial thermal velocity. To enforce a
vanishing initial total momentum, the velocity of every particle is
subsequently subtracted by the amount $N^{-1}\sum_i \mathbf{v}_i(0)$.

The velocities are updated from time $t$ to time
$t+h$, where the time step $h$ is much smaller than the mean
free time, by following two successive stages: collisions and free
streaming. In the collision stage, a sample of
$\frac{1}{2}N\omega_{\mbox{\scriptsize{max}}}h$ pairs is chosen at random
with equiprobability, where $\omega_{\mbox{\scriptsize{max}}}$ is an upper
bound estimate of the probability that a particle collides per unit of time.
For each pair $ij$ belonging to this sample, the following steps are taken:
(1) a given direction $\widehat{\boldsymbol{\sigma}}_{ij}$ is chosen at
random with equiprobability; (2) the collision between particles $i$ and $j$
is accepted with a probability equal to
$\Theta(\mathbf{v}_{ij}\cdot\widehat{\boldsymbol{\sigma}}_{ij})
\omega_{ij}/
\omega_{\mbox{\scriptsize{max}}}$,
where $\omega_{ij}=(4\pi\sigma^2\chi n)
|\mathbf{v}_{ij}\cdot\widehat{\boldsymbol{\sigma}}_{ij}|$; if the collision
is accepted, postcollisional velocities are assigned to both particles:
$\mathbf{v}_{i,j}\to\mathbf{v}_{i,j}\mp\frac{1}{2}(1+\alpha)
(\mathbf{v}_{ij}\cdot\widehat{\boldsymbol{\sigma}}_{ij})
\widehat{\boldsymbol{\sigma}}_{ij}$.
In the case that in one of the collisions $\omega_{ij}>
\omega_{\mbox{\scriptsize{max}}}$, the estimate of
$\omega_{\mbox{\scriptsize{max}}}$ is updated as
$\omega_{\mbox{\scriptsize{max}}}=\omega_{ij}$.

In the free streaming stage the velocity of every particle is changed
according to the thermostat force under consideration:
\begin{equation}
\label{54}
\mathbf{v}_i\to \mathbf{v}_i +\mathbf{w}_i,
\end{equation}
where
\begin{equation}
\label{55}
\mathbf{w}_i=\frac{1}{m}\int_t^{t+h}dt' \,
\mathbf{F}_i^{\mbox{\scriptsize{th}}}(t').
\end{equation}
In the case of the stochastic thermostat, Eq.\ (\ref{7}), one has
\begin{equation}
\label{56}
\langle
\mathbf{w}_i\rangle =\mathbf{0},\quad
\langle
\mathbf{w}_i
\mathbf{w}_j\rangle
=\mathsf{I}\xi_0^2 h \delta_{ij}.
\end{equation}
Consequently, each vector $\mathbf{w}_i$ is randomly drawn from the Gaussian
probability distribution
\begin{equation}
\label{57}
P(\mathbf{w})=\left(2\pi\xi_0^2h\right)^{3/2}e^{-w^2/2\xi_0^2h}.
\end{equation}
In the case of deterministic external forces the velocity increment
$\mathbf{w}_i$ is assigned in a more direct way.
If the thermostat is the Gaussian one, Eq.\ (\ref{11}),
\begin{equation}
\label{58}
\mathbf{w}_i=\left(e^{\zeta h}-1\right) \mathbf{v}_i.
\end{equation}
In the case of the non-Gaussian thermostat defined by Eq.\ (\ref{13}),
\begin{equation}
\label{59}
\mathbf{w}_i=g h \left(\widehat{\mathbf{v}}_i-\mathbf{k}\right),
\end{equation}
where the vector $\mathbf{k}\equiv N^{-1}\sum_i \widehat{\mathbf{v}}_i$ is
introduced to preserve the detailed conservation of momentum, i.e. $\sum_i
\mathbf{w}_i=\mathbf{0}$.

The moments of the distribution are simply obtained as
\begin{equation}
\label{60}
\langle v^p\rangle =\frac{1}{N}\sum_{i=1}^N v_i^p,\quad
\langle c^p\rangle=\langle v^p\rangle/v_0^p,
\end{equation}
where $v_0=\left(2\langle v^2\rangle/3\right)^{1/2}$. The
evaluation of the collisional moments $\mu_p$, $p=2 \mbox{ and
}4$, is more complicated. In the Appendix it is shown that
\begin{equation}
\label{61}
\mu_p=n^{-2} v_0^{-p-1}\int d\mathbf{v}_1 \int d\mathbf{v}_2\,
f(\mathbf{v}_1)f(\mathbf{v}_2) \Phi_p(\mathbf{v}_{1},\mathbf{v}_{2}),
\end{equation}
where
\begin{equation}
\label{62}
\Phi_2(\mathbf{v}_{1},\mathbf{v}_{2})=\frac{\pi(1-\alpha^2)}{8}v_{12}^3,
\end{equation}
\begin{eqnarray}
\label{63}
\Phi_4(\mathbf{v}_{1},\mathbf{v}_{2})
&=&\frac{\pi}{4}v_{12}\left\{
\frac{5(1-\alpha^2)}{3}v_{12}^2V_{12}^2
\right.\nonumber\\
&&
+\frac{(1-\alpha^2)(2+\alpha^2)}{12}
v_{12}^4
\nonumber\\
&&\left.+
(3-\alpha)(1+\alpha) \left[
\left(\mathbf{v}_{12}\cdot
\mathbf{V}_{12}\right)^2-\frac{1}{3}v_{12}^2V_{12}^2\right]\right\}.\nonumber\\
&&
\end{eqnarray}
In the above equations,
$\mathbf{v}_{12}\equiv \mathbf{v}_{1}-\mathbf{v}_{2}$,
$\mathbf{V}_{12}\equiv \frac{1}{2}(\mathbf{v}_{1}+\mathbf{v}_{2})$.
Starting from the exact expression (\ref{61}) and using (\ref{53}), we
arrive at the following formula for the numerical computation of $\mu_p$:
\begin{equation}
\label{64}
\mu_p=v_0^{-p-1}
\frac{1}{N'}{\sum_{ij}}'
\Phi_p(\mathbf{v}_{i},\mathbf{v}_{j}).
\end{equation}
The prime in the summation means that we restrict ourselves to $N'$ pairs $ij$
randomly  chosen out of the total number $N(N-1)/2$ of pairs in the
system. This allows us to
compute $\langle c^p\rangle$ and $\mu_p$ with similar accuracy within
reasonable computer times.
Once the steady state is reached, the relevant quantities
are subsequently averaged over $M$ independent instantaneous values.

In our simulations we have typically  taken
  $N=2\times 10^{5}$, $N'= 10^{7}$ and $M=10^3$.
Since the thermal velocity is not constant in the transient regime, we have
taken a time-dependent time step $h=0.01\lambda/v_0(t)$, where
$\lambda=(\sqrt{2}\pi\chi n\sigma^2)^{-1}$ is the mean free path.

\section{Results}
\label{sec:4}
By using the numerical method described in the previous section, we have
computed the steady-state values of the first few moments $\langle
c^p\rangle$ and $\mu_p$. We have also evaluated  the reduced
velocity distribution function $\widetilde{f}(\mathbf{c})$.
As a test of the accuracy of the simulations and also to check that the
steady state has been reached, we compare in Table \ref{tab:1}
the values of $\mu_4$ obtained
directly from Eq.\ (\ref{64}) with those
given by Eqs.\ (\ref{22}), (\ref{37}) or
(\ref{43bis}).
The values
corresponding to a Maxwell-Boltzmann distribution, $\mu_4^{(0)}$, are also
included in the table.
We can observe that the direct and indirect routes to the computation of
$\mu_4$ disagree less than $0.1\%$ in all the cases.
The difference between $\mu_4$ and $\mu_4^{(0)}$ is a measure of the
departure of $\widetilde{f}(\mathbf{c})$ from $\phi(\mathbf{c})$.

\begin{table*}
\caption{Comparison of the values of $\mu_4$ obtained directly with
those obtained indirectly from $\mu_2$.}
\label{tab:1}
\begin{tabular}{cccccccc}
\hline\noalign{\smallskip}
 & &\multicolumn{2}{c}{Stochastic}&\multicolumn{2}{c}{Gaussian}
 &\multicolumn{2}{c}{non-Gaussian}\\
 \cline{3-8}
$\alpha$ & $\mu_4^{(0)}$ &$\mu_4$&Eq.\ (\protect\ref{22})&
$\mu_4$&Eq.\ (\protect\ref{37})&
$\mu_4$&Eq.\ (\protect\ref{43bis})  \\
\noalign{\smallskip}\hline\noalign{\smallskip}
0.2 &10.925  &12.157 &12.155 &13.881 &13.881 &8.744&8.750  \\
0.4 &9.812  &10.602 &10.600 &11.494 &11.488 &7.631 &7.631  \\
0.6 &7.797  &8.036 &8.038 &8.213 &8.217 &5.811 &5.810  \\
0.8 &4.638  &4.499 &4.503 &4.414 &4.412 &3.335 &3.333  \\
\noalign{\smallskip}\hline
\end{tabular}
\end{table*}

Now we present the results separately for each one of the three thermostats
considered.
\subsection{Stochastic thermostat}
\label{sec:4.1}
\begin{figure}
\resizebox{\hsize}{!}{
  \includegraphics{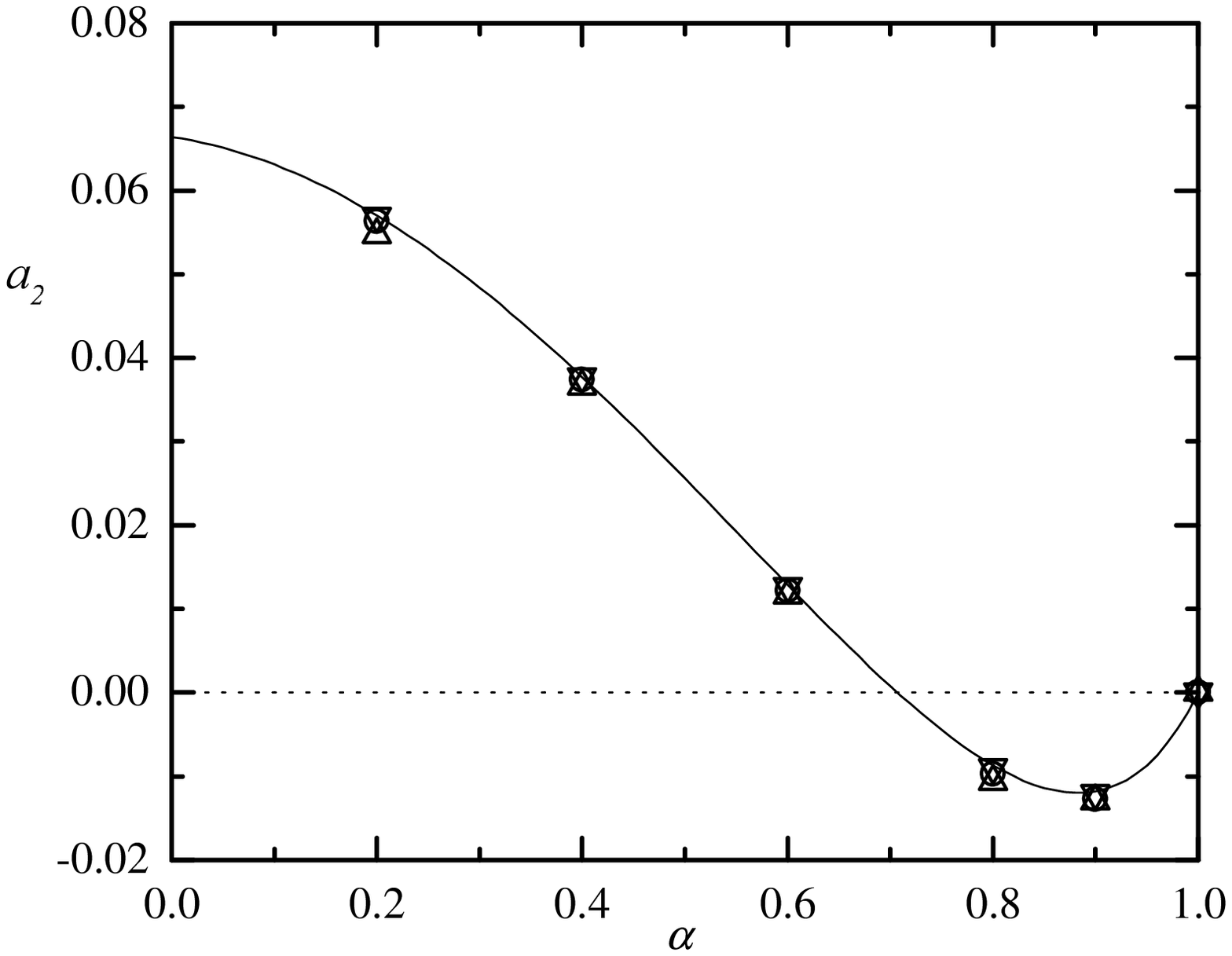}
}
\caption{Plot of the simulation values of $a_2$ ($\bigcirc$), $(\mu_2-
\mu_2^{(0)})/\mu_2^{(1)}$ ($\bigtriangleup$) and $(\mu_4-
\mu_4^{(0)})/\mu_4^{(1)}$ ($\bigtriangledown$) versus $\alpha$ in the
case of the stochastic thermostat. The solid line is the theoretical
estimate, Eq.\ (\protect\ref{9}).}
\label{fig:1}
\end{figure}

\begin{figure}
\resizebox{\hsize}{!}{
  \includegraphics{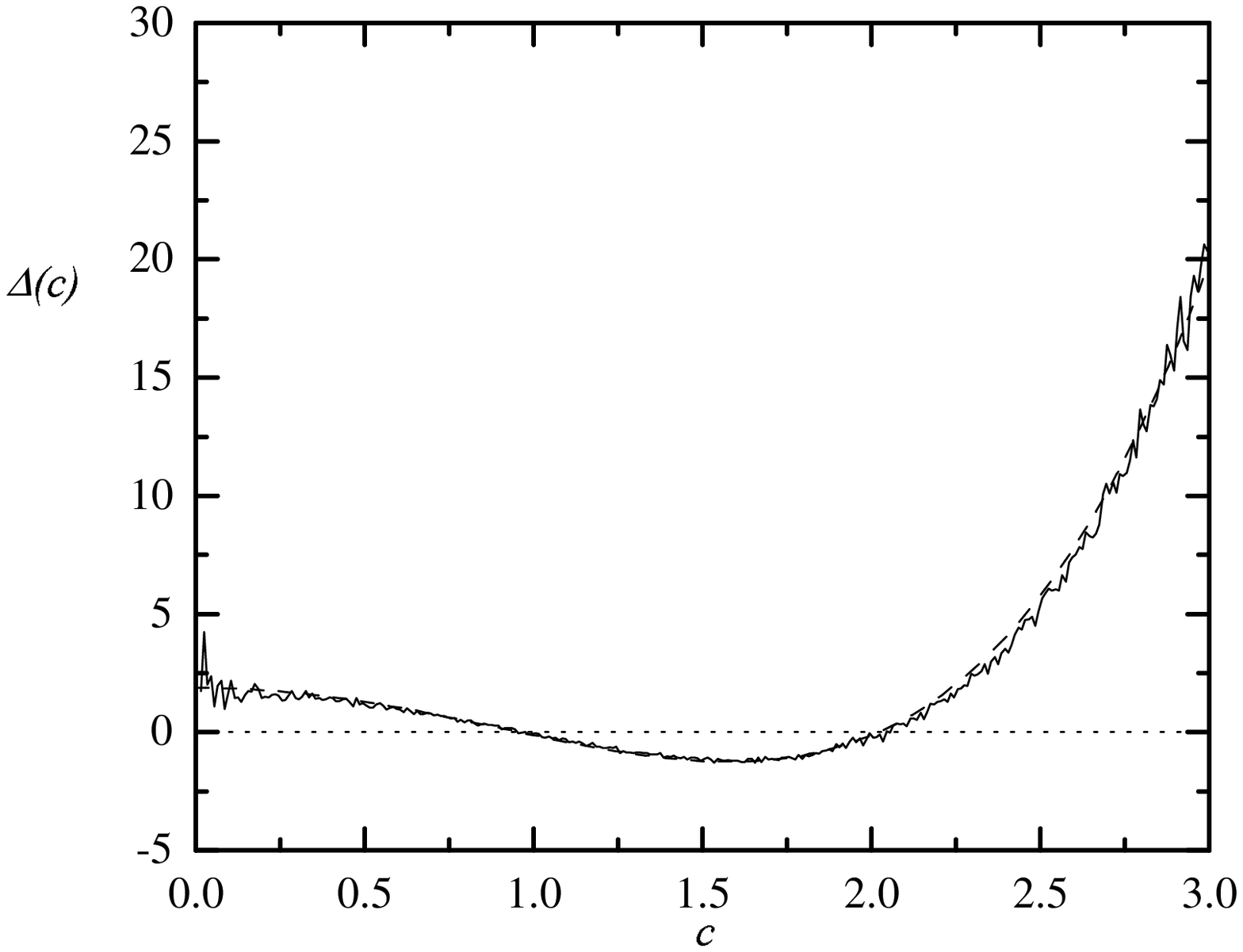}
}
\caption{Plot of the  simulation values of the function $\Delta(c)$ defined
by Eq.\ (\protect\ref{44}) for $\alpha=0.5$ in the case of the stochastic
thermostat. The dashed line is the Sonine polynomial $S_2(c^2)$. }
\label{fig:2}
\end{figure}

The basic quantity measuring the deviation of the distribution function from
the Maxwell-Boltzmann distribution is the cumulant $a_2$, Eq.\ (\ref{3}).
Figure \ref{fig:1} shows the $\alpha$-dependence of the simulation values of
$a_2$,  $(\mu_2- \mu_2^{(0)})/\mu_2^{(1)}$, $(\mu_4-
\mu_4^{(0)})/\mu_4^{(1)}$ and the theoretical estimate  (\ref{9}),
first derived in Ref.\ \cite{NE98}. As said in Sect.\ \ref{sec:2}, the
estimate (\ref{31}) gives practically the same results as (\ref{9}) and
therefore it is not plotted. The agreement between the simulation data and
the theoretical prediction is excellent, thus indicating
that the approximation (\ref{25}) was justified.
The above agreement indicates that the distribution function
$\widetilde{f}(\mathbf{c})$ for thermal velocities is well represented by
Eq.\ (\ref{23}). To confirm this,  the function
$\Delta(c)$ defined by Eq.\ (\ref{44}) is plotted in Fig.\ \ref{fig:2} for
$\alpha=0.5$. The simulation curve agrees very well with the Sonine
polynomial $S_2(c^2)$.
It is worth noting that this deviation from the Maxwell-Boltzmann distribution
in the case of the
stochastic thermostat could not be observed in recent two-dimensional
molecular dynamics simulations \cite{PO98} because the  statistical accuracy was not high enough.

\begin{figure}
\resizebox{\hsize}{!}{
  \includegraphics{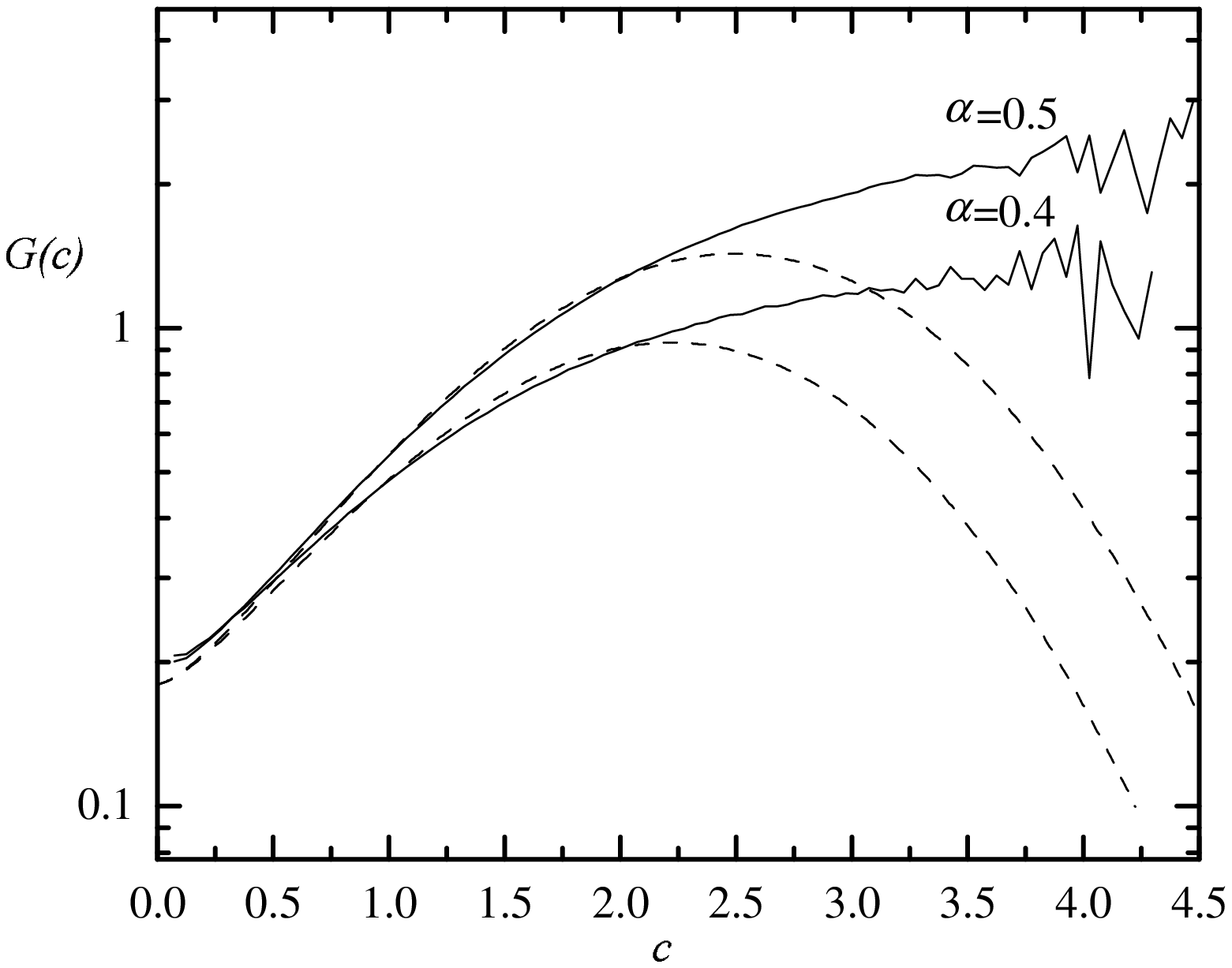}
}
\caption{Plot of the  simulation values of the function $G(c)$ defined
by Eq.\ (\protect\ref{66}) for $\alpha=0.4$ and $\alpha=0.5$ in the case of
the stochastic thermostat. The dashed lines
are the Maxwell-Boltzmann predictions.}
\label{fig:3}
\end{figure}

The theoretical prediction for the asymptotic high energy tail, Eq.\
(\ref{34}), is much harder to confirm in the simulations since it involves a
very small fraction of particles. Equation (\ref{34}) implies that
\begin{equation}
\label{65}
\lim_{c\to\infty} G(c)=K=\mbox{const},
\end{equation}
where
\begin{equation}
\label{66}
G(c)\equiv e^{A c^{3/2}}\widetilde{f}(\mathbf{c}).
\end{equation}
The function $G(c)$ is plotted (in logarithmic scale) in Fig.\ \ref{fig:3}
for $\alpha=0.4$ and $\alpha=0.5$. In both cases the values of $A$ have been
obtained from (\ref{34}) by using the simulation values of $\mu_2$, which
yields $A\simeq 1.99$ ($\alpha=0.4$) and $A\simeq 2.11$ ($\alpha=0.5$). The
figure is convincingly consistent with Eq.\ (\ref{65}), where $K\simeq 1.3$
and $K\simeq 2.2$  for $\alpha=0.4$ and $\alpha=0.5$, respectively.
Figure \ref{fig:3} also shows the corresponding functions $G(c)$ obtained
from Eq.\ (\ref{66}) by replacing $\widetilde{f}(\mathbf{c})$ by the
Maxwell-Boltzmann distribution $\phi(\mathbf{c})$. The overpopulation
phenomenon for $c>2$ is quite apparent. At $c=4$, for instance,
$\widetilde{f}/\phi\simeq 8$ for $\alpha=0.4$ and
$\widetilde{f}/\phi\simeq 5$ for $\alpha=0.5$.

\subsection{Gaussian thermostat}
\label{sec:4.2}
\begin{figure}
\resizebox{\hsize}{!}{
  \includegraphics{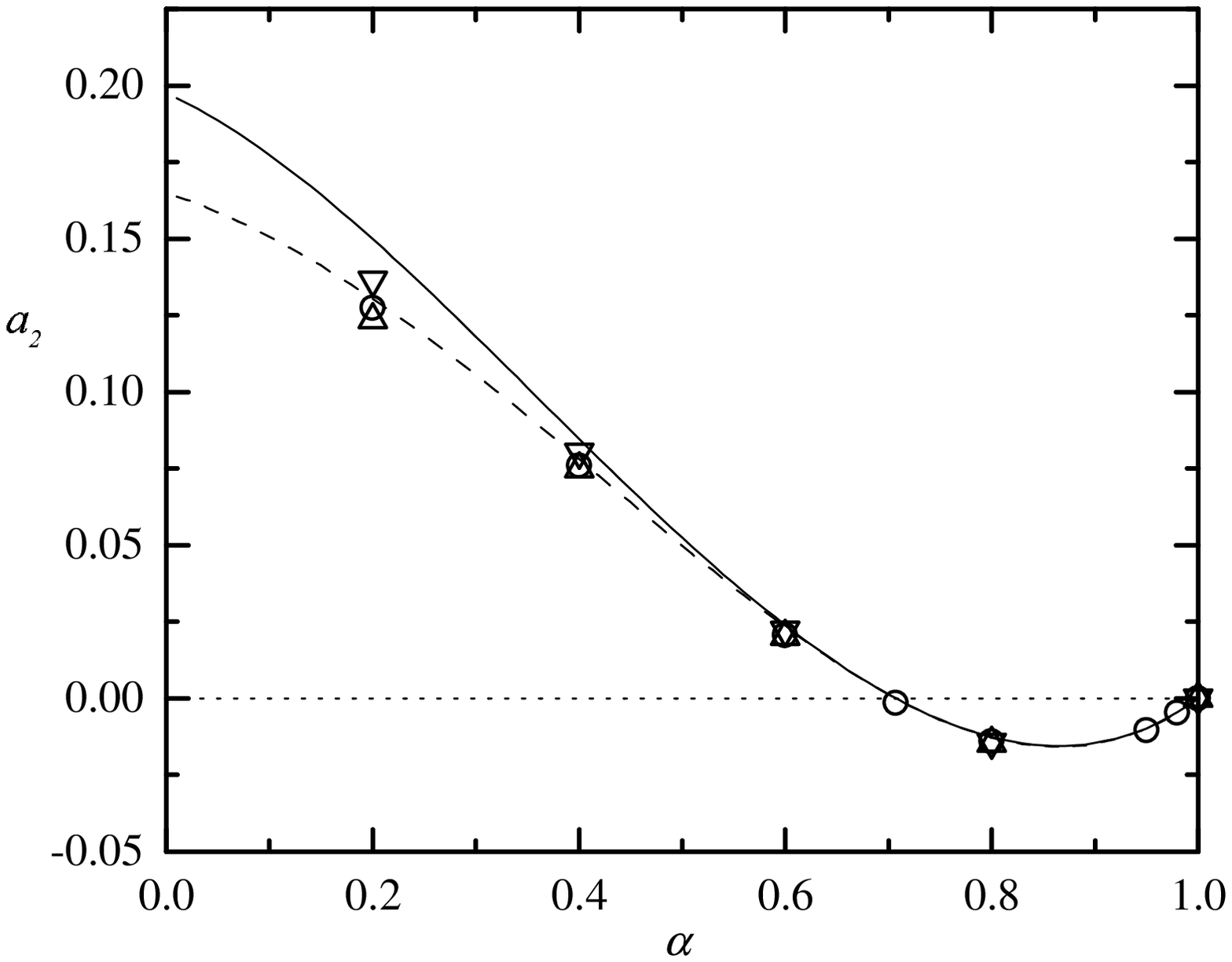}
}
\caption{Plot of the simulation values of $a_2$ ($\bigcirc$), $(\mu_2-
\mu_2^{(0)})/\mu_2^{(1)}$ ($\bigtriangleup$) and $(\mu_4-
\mu_4^{(0)})/\mu_4^{(1)}$ ($\bigtriangledown$) versus $\alpha$ in the
case of the Gaussian thermostat. The solid and dashed lines are the
theoretical estimates\ (\protect\ref{5}) and (\protect\ref{39}),
respectively.}
\label{fig:4}
\end{figure}

\begin{figure}
\resizebox{\hsize}{!}{
  \includegraphics{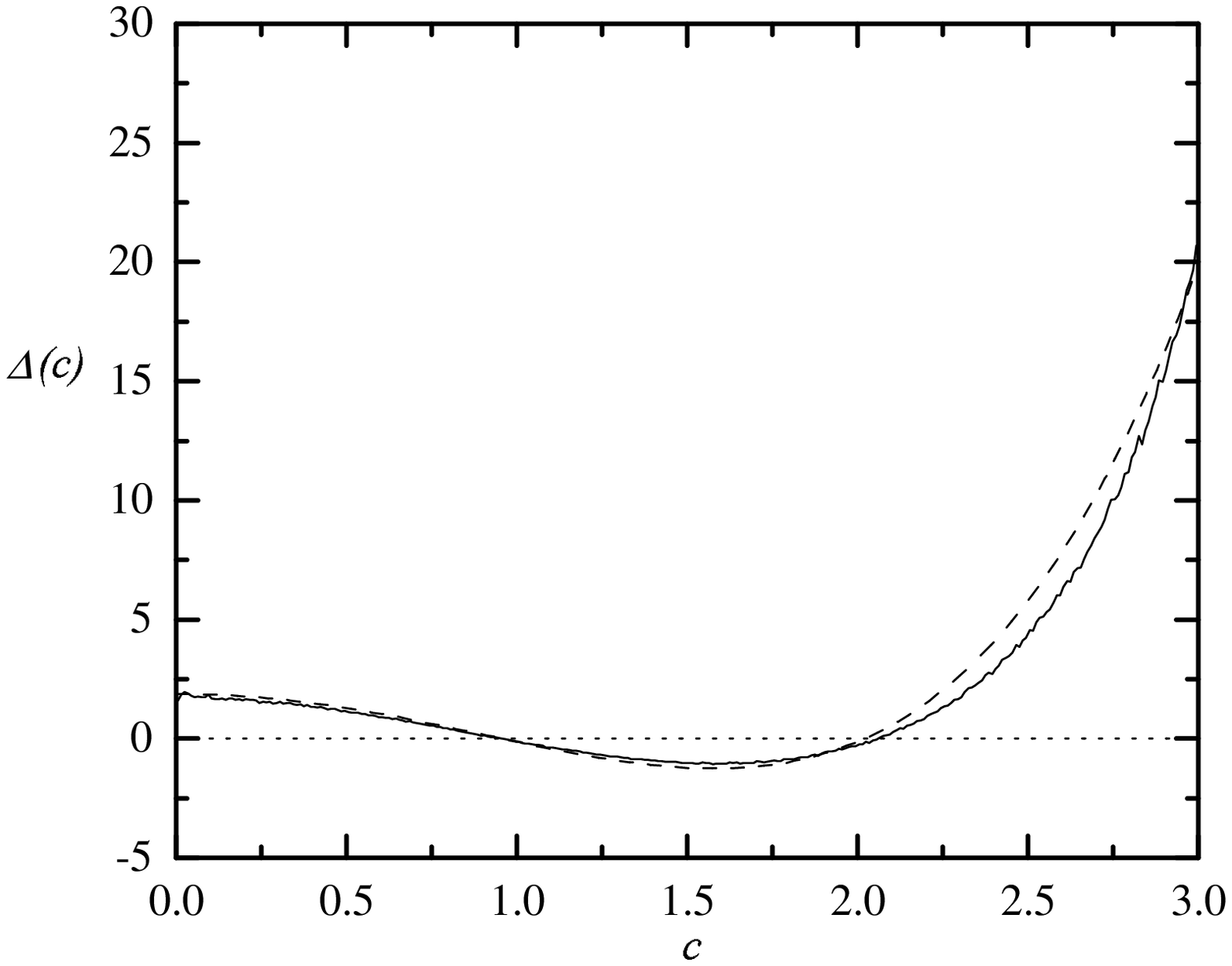}
}
\caption{Plot of the  simulation values of the function $\Delta(c)$ defined
by Eq.\ (\protect\ref{44}) for $\alpha=0.4$ in the case of the Gaussian
thermostat. The dashed line is the Sonine polynomial $S_2(c^2)$. }
\label{fig:5}
\end{figure}

\begin{figure}
\resizebox{\hsize}{!}{
  \includegraphics{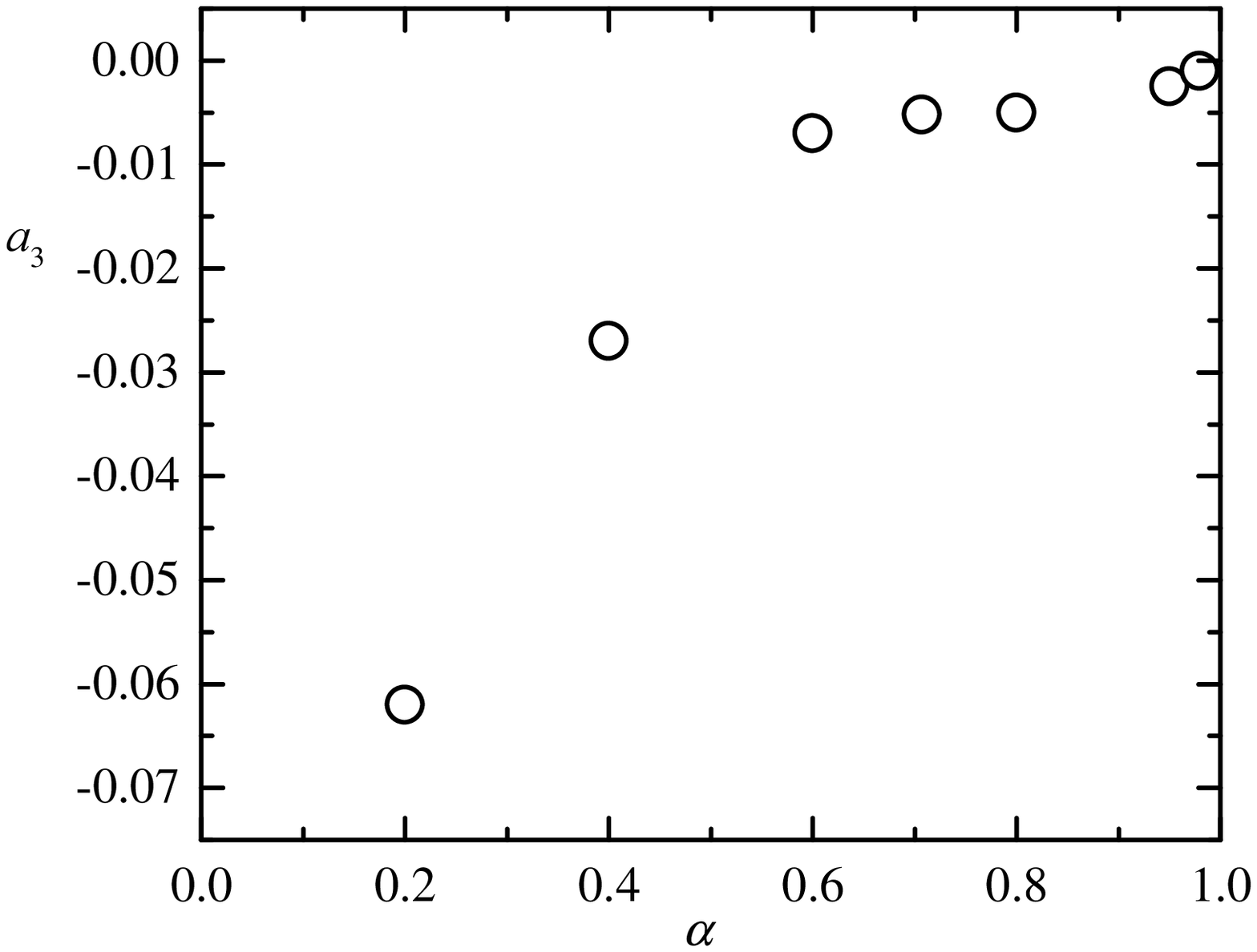}
}
\caption{Plot of the simulation values of $a_3$  versus $\alpha$ in the case
of the Gaussian thermostat. }
\label{fig:10}
\end{figure}

Now we carry out a parallel analysis in the case of the deterministic
Gaussian thermostat.
The $\alpha$-dependence of the simulation values of
$a_2$,  $(\mu_2- \mu_2^{(0)})/\mu_2^{(1)}$ and $(\mu_4-
\mu_4^{(0)})/\mu_4^{(1)}$  are shown in Figure \ref{fig:4}.
The values of $a_2$ are in this case generally larger than in the previous
case. In addition, Eq.\ (\ref{25}) tends to overestimate $\mu_4$ and
underestimate $\mu_2$ for small $\alpha$.
As a consequence, the theoretical estimate (\ref{5}) gives values larger
than the simulation data for $\alpha<0.5$, while the estimate (\ref{39}) is
fairly good in that region.
For values of the coefficient of restitution for which the fourth cumulant
$a_2$ is not small enough (say $a_2\simeq 0.1$), we may  expect a
non-negligible deviation from (\ref{23}). This is confirmed in Fig.\
\ref{fig:5}, where $\Delta (c)$ is plotted for $\alpha=0.4$. Here the
contributions associated with higher-order Sonine polynomials are relatively
important.
As a quantitative measure of the difference between $\Delta(c)$ and
$S_2(c^2)$,
we have obtained preliminary simulation results for the
sixth  cumulant $a_3$ defined as
\begin{eqnarray}
\label{67bis}
a_3&\equiv &\frac{48}{d(d+2)(d+4)}\int d\mathbf{c} \,S_3(c^2)
\widetilde{f}(\mathbf{c})
\nonumber\\
&=&-\frac{8}{d(d+2)(d+4)}\langle c^6\rangle +1+3a_2.
\end{eqnarray}
This quantity is plotted in Fig.\ \ref{fig:10} for $d=3$.
For $\alpha\geq 0.6$ $|a_3|$ remains small, but for larger dissipation the
values of $|a_3|$ increase rapidly.

The high energy tail predicted by Eq.\ ({\ref{41}) \cite{NE98,EP97} is
tested in Fig.\ \ref{fig:6}, where
\begin{equation}
\label{67}
G(c)\equiv e^{A c}\widetilde{f}(\mathbf{c})
\end{equation}
is plotted for $\alpha=0.2$ and $\alpha=0.4$. The corresponding values of
$A$ are $A\simeq 3.82$ and $A\simeq 4.41$, respectively.
The agreement with Eq.\ (\ref{65}) is excellent; from the simulation data
we can estimate  $K\simeq 7$ for $\alpha=0.2$ and $K\simeq 31$ for
$\alpha=0.4$. \begin{figure}
\resizebox{\hsize}{!}{
  \includegraphics{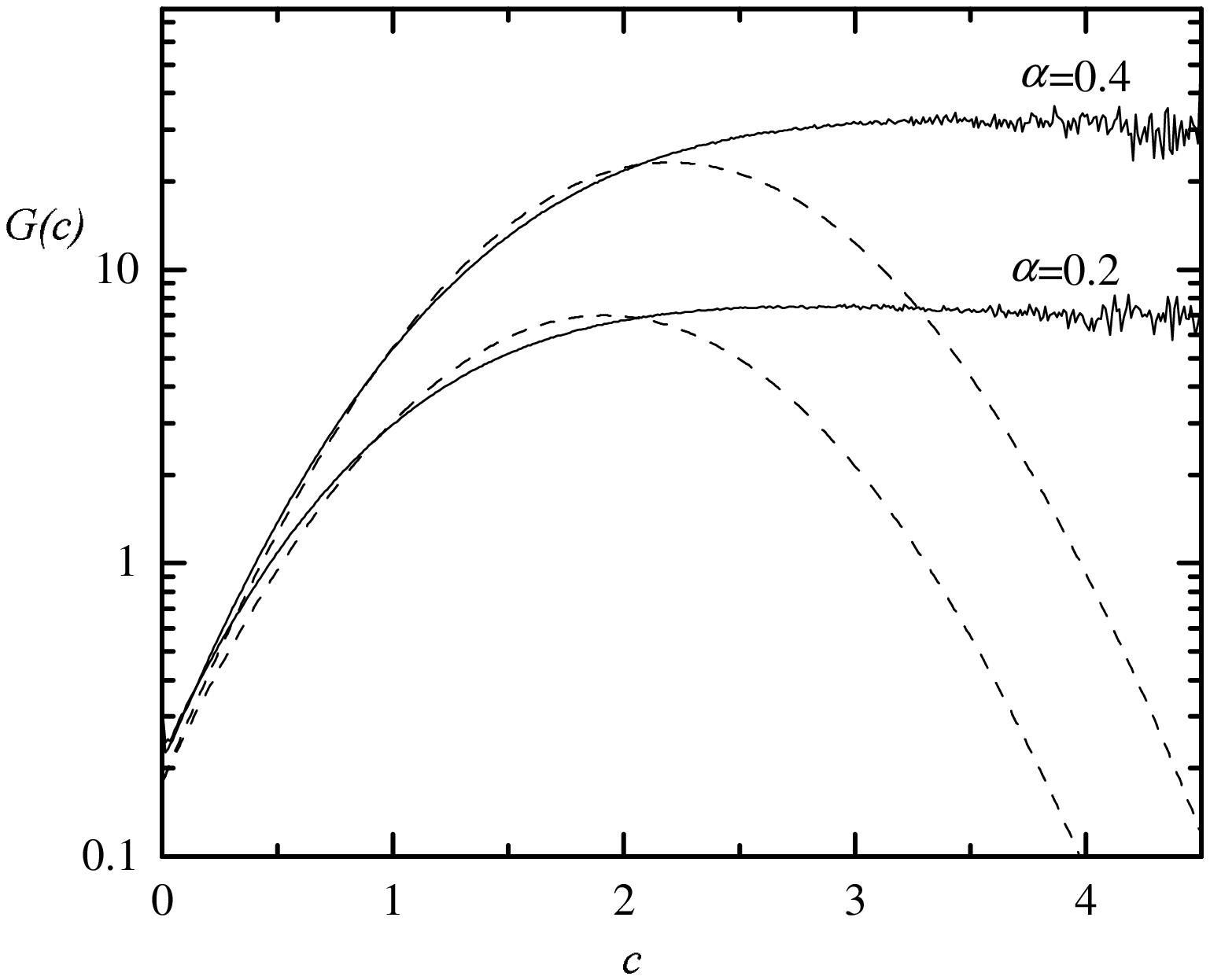}
}
\caption{Plot of the  simulation values of the function $G(c)$ defined
by Eq.\ (\protect\ref{67}) for $\alpha=0.2$ and $\alpha=0.4$ in the case of
the Gaussian thermostat. The dashed lines
are the Maxwell-Boltzmann predictions.}
\label{fig:6}
\end{figure}
In this case of a Gaussian thermostat, the overpopulation effect is
much more important than in the previous case.
At $c=4$,
$\widetilde{f}/\phi\simeq 80$ for $\alpha=0.2$ and
$\widetilde{f}/\phi\simeq 34$ for $\alpha=0.4$.
The results reported here for inelastic hard spheres complement those obtained
by Brey et al.\ \cite{BCR99}, where the asymptotic
behavior (\ref{41}) was verified  for inelastic hard disks.

Recently, Sela and Goldhirsch \cite{SG98} have obtained \textit{numerically}
the function $\Delta(c)$ in the \textit{low dissipation} limit. In their notation,
$\lim_{\alpha\to 1}\Delta(c)\equiv -8\widehat{\Phi}_\epsilon(c)$. {}From
simulation results presented in Ref.\ \cite{BMC96} for
$\alpha=0.99$ it follows that the function  $\widehat{\Phi}_\epsilon(c)$ is
well represented by the Sonine polynomial $-S_2(c^2)/8$ in the range
$0\leq c\leq 1$. However,
this agrees only qualitatively with the function  $\widehat{\Phi}_\epsilon(c)$
obtained numerically by Sela and Goldhirsch \cite{SG98}. For instance, from Fig.\ 3
of Ref.\ \cite{SG98} one  gets $\widehat{\Phi}_\epsilon(0)\simeq -0.35$,
while  $-S_2(0)/8=-15/64\simeq -0.23$.  Moreover, it is claimed in Ref.\
\cite{SG98} that $\widehat{\Phi}_\epsilon(c)\sim c^2\log c$ for large $c$, which
differs from the behavior (\ref{41}) that has been confirmed here and in Ref.\
\cite{BCR99}.  It is possible that the high energy tail obtained
from the perturbative approach presented in Ref.\
\cite{SG98}  only holds for $1\ll c\ll
(1-\alpha^2)^{-1}$ and  thus it is not representative of the general
asymptotic behavior for arbitrary $\alpha$.

\subsection{Non-Gaussian thermostat}
\label{sec:4.3}
In contrast to the two previous cases, the Sonine polynomials $\{S_p(c^2)\}$
are not expected to constitute a good set for the expansion of the ratio
$\widetilde{f}(\mathbf{c})/\phi(\mathbf{c})$ in the case of the
non-Gaussian thermostat (\ref{14}) since the latter is singular at
$\mathbf{c}=\mathbf{0}$. Consequently, we do not expect the estimate
(\ref{48}) to be quantitatively accurate. This is confirmed in
Fig.\ \ref{fig:7}, where we observe that Eq.\ (\ref{48}) gives values that
are about 20\% smaller in magnitude than the simulation ones.
\begin{figure}
\resizebox{\hsize}{!}{
  \includegraphics{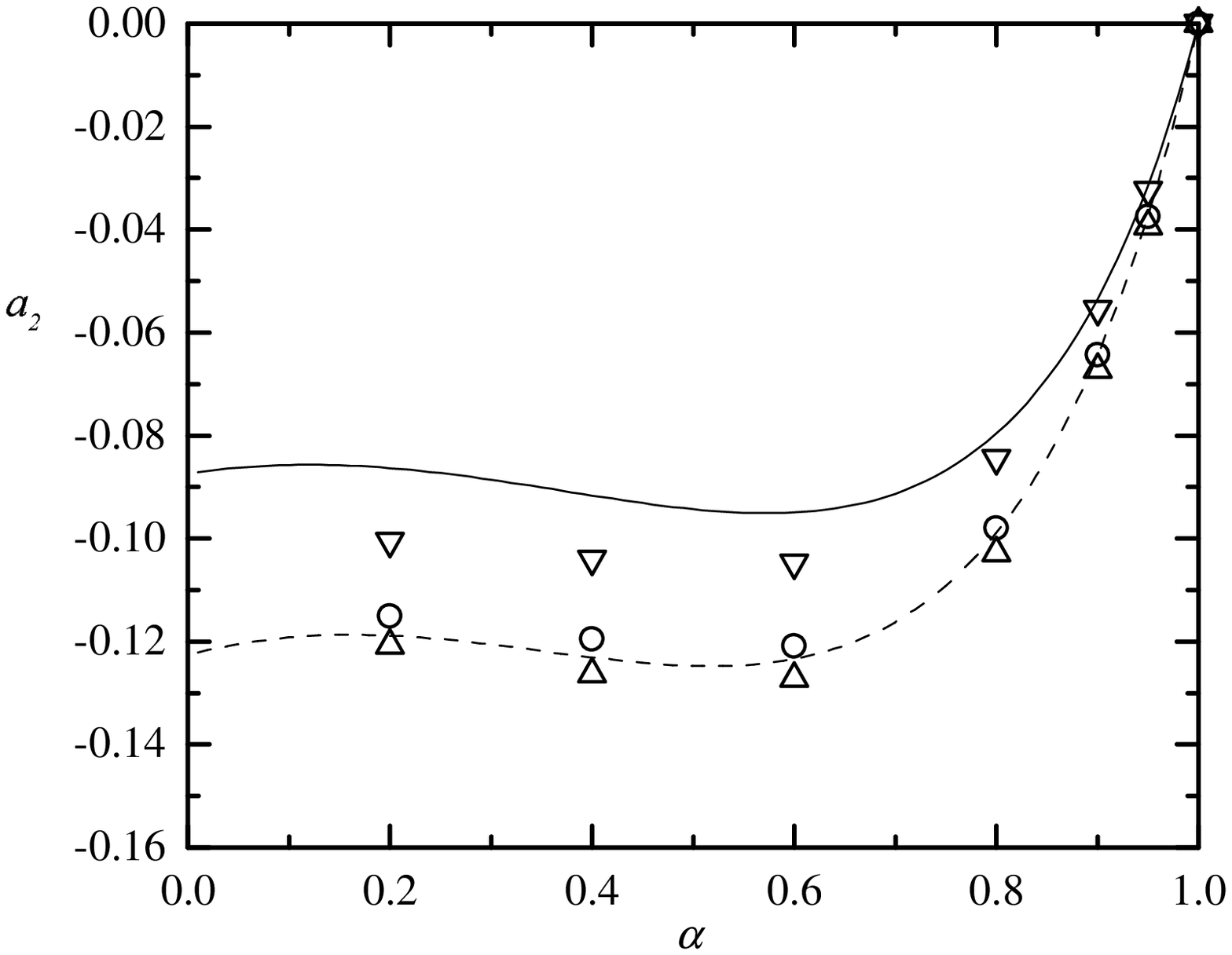}
}
\caption{Plot of the simulation values of $a_2$ ($\bigcirc$), $(\mu_2-
\mu_2^{(0)})/\mu_2^{(1)}$ ($\bigtriangleup$) and $(\mu_4-
\mu_4^{(0)})/\mu_4^{(1)}$ ($\bigtriangledown$) versus $\alpha$ in the
case of the non-Gaussian thermostat. The solid and dashed lines are the
estimates (\protect\ref{48}) and (\protect\ref{68}),
respectively.}
\label{fig:7}
\end{figure}
Also, the
approximation (\ref{25}) with $\mu_{2,4}^{(1)}$ given by Eqs.\ (\ref{27})
and (\ref{29})  is rather poor. It is
reasonable to expect that a better approximation would be obtained if
$\langle c\rangle$, $\langle c^3\rangle$, $\mu_{2}$ and $\mu_{4}$ were
computed from the unknown function $\Delta(c)$ rather than from $S_2(c^2)$.
When plotting the simulation data of
$(\mu_{2}-\mu_{2}^{(0)})/\mu_{2}^{(1)}$,
$(\mu_{4}-\mu_{4}^{(0)})/\mu_{4}^{(1)}$,
$\langle c\rangle$ and $\langle c^3\rangle$
 versus $a_2$ we have
observed that the points fit well in straight lines, as predicted by Eqs.\
 (\ref{25}), (\ref{47}) and (\ref{47bis}), but with different slopes. More
specifically, our simulation results indicate that, instead of Eqs.\
(\ref{25}), (\ref{47}) and (\ref{47bis}), one should have
 (for $d=3$)
\begin{equation}
\label{67.2}
\mu_2\simeq\mu_2^{(0)}+\frac{21}{20}\mu_2^{(1)}a_2,
\end{equation}
\begin{equation}
\label{67.3}
\mu_4\simeq\mu_4^{(0)}+\frac{13}{15}\mu_4^{(1)}a_2,
\end{equation}
\begin{equation}
\label{67.0}
\langle c\rangle\simeq \frac{2}{\pi^{1/2}}
\left(1-\frac{3}{2}\frac{1}{8}a_2\right),
\end{equation}
\begin{equation}
\label{67.1}
\langle c^3\rangle\simeq
\frac{2\sqrt{2}}{\pi^{1/2}}
\left(1+\frac{23}{20}\frac{3}{8}a_2\right).
\end{equation}
If we insert the above expressions into Eq.\  (\ref{43bis}) and neglect
terms nonlinear in $a_2$, we get
\begin{equation}
\label{68}
a_2\simeq
-\frac{240(1-\alpha)(1+2\alpha^2)}{1957-1125\alpha+390(1-\alpha)\alpha^2}.
\end{equation}
This semi-empirical estimate exhibits a fairly good agreement with the
simulation data, as shown in Fig.\ \ref{fig:7}.

The limitations of a Sonine description in the case of the non-Gaussian
thermostat are quite apparent in
 Fig.\ \ref{fig:8}, where $\Delta (c)$ is
plotted for $\alpha=0.4$, 0.6 and 0.95.
The curves corresponding to $\alpha=0.4$ and  $\alpha=0.6$ practically
coincide, while the curve corresponding to $\alpha=0.95$ clearly deviates in
the region of very small velocities. As a matter of fact, $\Delta(0)$ is
roughly equal to $-a_2^{-1}$, which indicates an almost vanishing population of rest
particles, i.e. $\widetilde{f}(\mathbf{0})\approx 0$, even at $\alpha=0.95$.
\begin{figure}
\resizebox{\hsize}{!}{
  \includegraphics{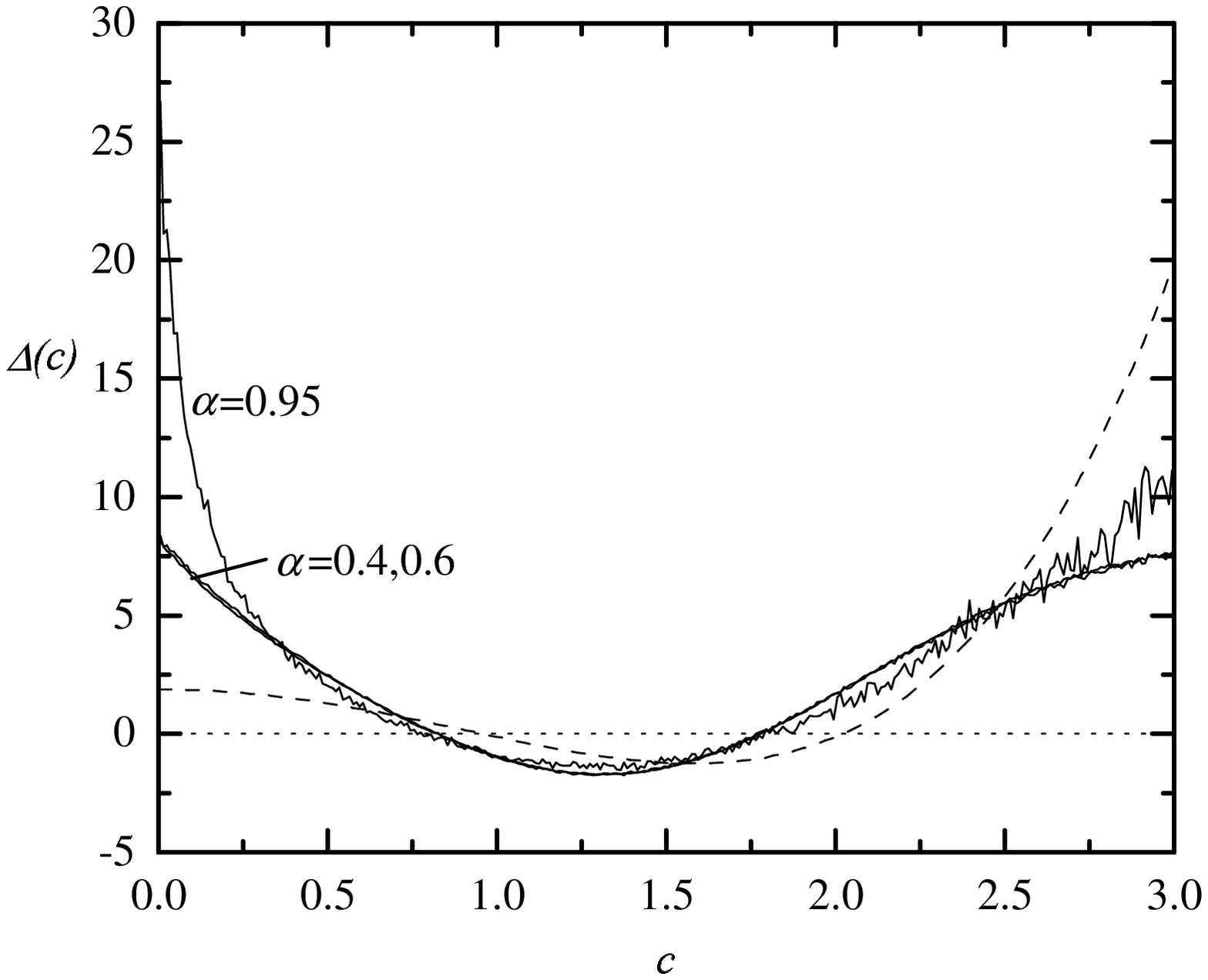}
}
\caption{Plot of the  simulation values of the function $\Delta(c)$ defined
by Eq.\ (\protect\ref{44}) for $\alpha=0.4$, 0.6 and 0.95
in the case of the non-Gaussian thermostat. The dashed line is the Sonine
polynomial $S_2(c^2)$. }
\label{fig:8}
\end{figure}
A key feature of Fig.\ \ref{fig:8} is the existence of a non-zero initial
slope, $\left. \partial \Delta(c)/\partial c\right|_{c=0}\neq 0$, that
cannot be described by any polynomial in $c^2$.

{}From the analysis made at the end of Subsect.\ \ref{sec:2.3}, we expect an
underpopulated high energy tail of the form (\ref{52}), where the
coefficient $A$ is unknown. By a fitting of the simulation results we have
estimated $A\simeq 1.48$ for $\alpha=0.3$ and $A\simeq 1.51$ for
$\alpha=0.4$.
Figure \ref{fig:9} shows the function
\begin{equation}
\label{69}
G(c)\equiv e^{A c^2}\widetilde{f}(\mathbf{c})
\end{equation}
for $\alpha=0.3$ and $\alpha=0.4$. In both cases the value of $K$ is
$K\simeq 1.7$.
 \begin{figure}
\resizebox{\hsize}{!}{
  \includegraphics{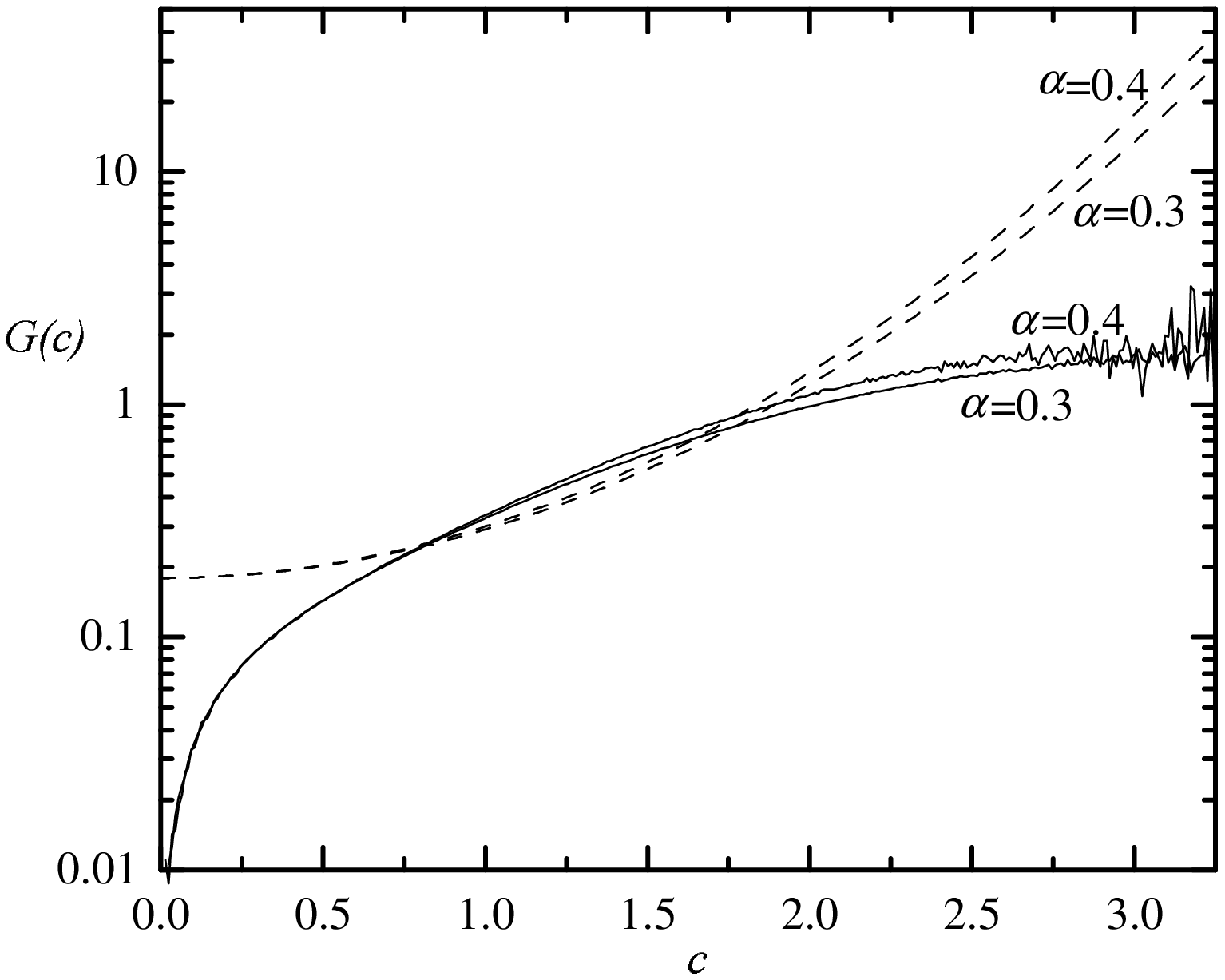}
}
\caption{Plot of the  simulation values of the function $G(c)$ defined
by Eq.\ (\protect\ref{69}) for $\alpha=0.3$ and $\alpha=0.4$ in the case of
the non-Gaussian thermostat. The dashed lines
are the Maxwell-Boltzmann predictions.}
\label{fig:9}
\end{figure}
The regions of small and large velocities are highly underpopulated with
respect to the Maxwell-Boltzmann distribution.
At $c=3$, for instance,
$\widetilde{f}/\phi\simeq 0.12$ for $\alpha=0.3$ and
$\widetilde{f}/\phi\simeq 0.10$ for $\alpha=0.4$.

\section{Summary and discussion}
\label{sec:5}
In this paper we have performed direct Monte Carlo simulations of the
Enskog-Boltzmann equation for a fluid of smooth inelastic spheres in
spatially uniform states.
Upon describing the velocity distribution of the granular fluid by  the
Enskog-Boltzmann equation (\ref{1}) it has been implicitly assumed the
validity of the ``molecular chaos'' hypothesis of uncorrelated binary
collisions. However, molecular dynamics simulations of hard disks have shown
 a non-uniform distribution of impact parameters for
high enough dissipation ($\alpha<0.8$) \cite{Luding}. In addition, there exist long range
spatial correlations in density and flow fields which cannot be understood
on the basis of the Enskog-Boltzmann equation \cite{NEBO97}. These two
effects are associated with the appearance of the so-called cluster
instability \cite{BM90} for systems sufficiently large. Since we have
simulated directly the spatially uniform equation (\ref{1}), such an
instability is precluded in the simulations.

To compensate for cooling
effects associated with the inelasticity of collisions, three types of
``thermostatting'' external driving forces have been considered. We have
analyzed the deviation of the steady-state velocity distribution function
from the Maxwell-Boltzmann distribution, as measured by the fourth cumulant
$a_2$ and by the high energy tail.

A simple mechanism for thermostatting the system is to assume that the
particles are subjected to random kicks \cite{WM96}, what  mimics the
effects of shaking or vibrating the vessel \cite{ER89}. If this stochastic
force has the properties of a white noise [cf. Eq.\ (\ref{7})], it gives
rise to a Fokker-Planck diffusion term in the Enskog-Boltzmann equation
\cite{NE98}. By making a first Sonine approximation, van Noije and Ernst
\cite{NE98} have obtained an approximate expression for $a_2$ as a function
of the coefficient of normal restitution $\alpha$. Our simulation results
confirm the accuracy of that expression even for  large dissipation
($\alpha=0.2$). We have also confirmed a high energy tail of the form
$f(\mathbf{v})\sim \exp[-A(v/v_0)^{3/2}]$ (where $v_0$ is the thermal
velocity) derived in Ref.\
\cite{NE98}. Moreover, that asymptotic behavior (which represents an
overpopulation with respect to the Maxwell-Boltzmann distribution) is
already practically reached for $v>4v_0$, at least for $\alpha=0.4
\mbox{ and }0.5$.

In the absence of any external forcing, the freely evolving granular fluid
reaches a  homogeneous cooling state in which all the time dependence of the
velocity distribution occurs through the thermal velocity $v_0(t)$, so that
the distribution of the \textit{reduced\/} velocity
$\mathbf{c}=\mathbf{v}/v_0$ is stationary. When the Enskog-Boltzmann
equation is written in terms of this reduced velocity, the operator
$\partial/\partial t$ gives rise to an operator that coincides
with the one representing the action of an external force proportional to
the particle velocity [cf. Eq.\ (\ref{11})]. This type of ``anti-drag''
force can also be
justified by Gauss's principle of least constraint \cite{EM90} and has been
widely used in nonequilibrium molecular dynamics simulations of molecular
fluids. Thus, the homogeneous  cooling state is equivalent to the steady
state reached under a Gaussian thermostat.
 In their simulations, Brey et al.\ \cite{BMC96,BCR99} used the former point of
 view,
while in this paper we have used the latter. Our simulations complement
those of Ref.\ \cite{BMC96} also in that we have considered a wide range
$0.2\leq \alpha\leq 1$, while Brey et al.\ \cite{BMC96} analyzed in detail
the region $0.7\leq \alpha\leq 1$. They obtained an excellent agreement with
the estimate (\ref{5}) based on a Sonine approximation, first derived in
Ref.\ \cite{NE96}. However, as $\alpha$ decreases and $a_2$ becomes larger,
we have seen in this paper that Eq.\ (\ref{5}) overestimates $a_2$.
This discrepancy can be traced back to contributions associated with
higher order Sonine polynomials as well as to the ambiguity
involved in the approximate determination of $a_2$  by neglecting nonlinear
terms in the exact equation (\ref{37}). If Eq.\  (\ref{37}) is rewritten in
another equivalent form (for example, by transferring a quantity from one
side to the other), the same method yields a different approximation for
$a_2$. As long as $a_2$ remains small (say $|a_2|<0.05$), all the
approximations give practically undistinguishable results. On the other
hand, for larger values of $a_2$ (i.e., for $\alpha<0.5$) the result is
relatively dependent of the route followed. By starting from Eq.\ (\ref{37})
rewritten as $\mu_4/(1+a_2)=(d+2)\mu_2$, we have obtained the estimate
(\ref{39}), which is seen to agree fairly well with the simulation results
for the whole range of coefficients of restitution considered.
The asymptotic  analysis of the kinetic equation
predicts a high energy tail of the form \cite{NE98,EP97}
$f(\mathbf{v})\sim \exp[-A(v/v_0)]$, what represents an overpopulation
phenomenon stronger than in the previous case.
This behavior was already confirmed in Ref.\ \cite{BCR99} for $d=2$ and
has now been confirmed by our simulation results for $d=3$.

In the case of the Gaussian thermostat, the heating force points in the
motion direction and its magnitude is proportional to that of the particle
velocity. This is a very efficient thermostat because it gives more energy
to fast particles, which are the ones colliding more frequently.
In contrast, the stochastic thermostat adds a velocity increment per unit of
time that is random both in direction and in magnitude. This is why the high
energy population is larger with the Gaussian thermostat than with
the stochastic thermostat. Nevertheless, in both cases such a population is
larger than in the case of elastic particles at equilibrium. One could be
tempted to expect that this overpopulation is a common feature of heated
granular fluids, regardless of the mechanism of heating. Our third choice of
thermostat, Eq.\ (\ref{13}), proves that this is not the case. Like in the
case of the stochastic thermostat, the force is independent of the magnitude
of the particle velocity; like in the case of the Gaussian thermostat, the
force is deterministic and points in the motion direction.
The action of this third thermostat can be graphically described
by saying that,
between two successive collisions, a particle feels a ``pseudo-gravity''
field that makes it to ``fall'' along its motion direction.
With this choice
of a non-Gaussian deterministic thermostat, the  Sonine polynomials
$\{S_p(c^2)\}$ are not a good set to represent the ratio
$\widetilde{f}(\mathbf{c})/\phi(\mathbf{c})$, even for low dissipation. As a
consequence, the theoretical estimate of $a_2$ derived by assuming that
$[\widetilde{f}(\mathbf{c})/\phi(\mathbf{c})-1]/a_2\equiv \Delta(c)\simeq
S_2(c^2)$, while being qualitatively correct, is not quantitatively
accurate. We have not been able to get the functional form of $\Delta(c)$
in the limit of low dissipation. However, we have estimated  its
contributions to $\langle c\rangle$, $\langle c^3\rangle$, $\mu_2$ and
$\mu_4$ from the simulation data. This has allowed us to obtain an
approximate expression for $a_2$ that fits well the simulation results.
An interesting feature of the velocity distribution function in this case is
that it is highly underpopulated with respect to the Maxwell-Boltzmann
distribution \textit{both\/} for small and large velocities.
Between two successive collisions, every particle experiences a constant
tangential acceleration $g$. The total work done by this force is exactly
compensated by the total loss of energy through collisions, which are much
more frequent for fast particles  than for slow ones. Therefore, the
population of slow particles decreases because of the action of the external
force, while that of fast particles decreases because of the effect of
collisions. The high energy tail of the distribution function is of the form
$\widetilde{f}(\mathbf{c})\sim \exp(-Ac^2)$ with $A>1$. In this case the
gain and loss terms of the collision integral are comparable, so that the
dependence of $A$ on $\alpha$ is an open problem.
\begin{acknowledgement}
Partial support from the DGES (Spain) through grant No. PB97-1501 and from
the Junta de Extremadura--Fondo Social Europeo through grant No. IPR98C019
is gratefully acknowledged.
\end{acknowledgement}
 \appendix
\section{Collisional moments}
\label{appA}
In this Appendix we derive the expressions (\ref{61})--(\ref{63}). Starting
from Eq.\ (\ref{21}) and by a standard change of
variables, it is easy to get  \cite{NE98}
\begin{equation}
\label{A1}
\mu_p=\int d\mathbf{c}_1\int d\mathbf{c}_2\,
\widetilde{f}(\mathbf{c}_1)\widetilde{f}(\mathbf{c}_2)
\widetilde{\Phi}_p(\mathbf{c}_1,\mathbf{c}_2),
\end{equation}
where
\begin{equation}
\label{A1bis}
\widetilde{\Phi}_p(\mathbf{c}_1,\mathbf{c}_2)=\frac{1}{2}\int
d\widehat{\boldsymbol{\sigma}}\,
\Theta(\mathbf{c}_{12}\cdot\widehat{\boldsymbol{\sigma}})
(\mathbf{c}_{12}\cdot\widehat{\boldsymbol{\sigma}})
\left[c_1^p+c_2^p-{c_1'}^p-{c_2'}^p\right],
\end{equation}
with $\mathbf{c}_{1,2}'=\mathbf{c}_{1,2}\mp\frac{1}{2}(1+\alpha)
(\mathbf{c}_{12}\cdot\widehat{\boldsymbol{\sigma}})
\widehat{\boldsymbol{\sigma}}$.
In the cases $p=2$ and $p=4$ we have
\begin{equation}
\label{A2}
c_1^2+c_2^2-{c_1'}^2-{c_2'}^2=
\frac{1-\alpha^2}{2}(\mathbf{c}_{12}\cdot\widehat{\boldsymbol{\sigma}})^2,
\end{equation}
\begin{eqnarray}
\label{A2bis}
c_1^4+c_2^4-{c_1'}^4-{c_2'}^4&=&(1-\alpha^2)
(\mathbf{c}_{12}\cdot\widehat{\boldsymbol{\sigma}})^2
\left(\frac{c_{12}^2}{4}+C_{12}^2\right)
\nonumber\\
&&
-\frac{(1-\alpha^2)^2}{8}
(\mathbf{c}_{12}\cdot\widehat{\boldsymbol{\sigma}})^4
\nonumber\\
&&-2(1+\alpha)^2
(\mathbf{c}_{12}\cdot\widehat{\boldsymbol{\sigma}})^2
(\mathbf{C}_{12}\cdot\widehat{\boldsymbol{\sigma}})^2
\nonumber\\
&&+4(1+\alpha)(\mathbf{c}_{12}\cdot\widehat{\boldsymbol{\sigma}})
(\mathbf{C}_{12}\cdot\widehat{\boldsymbol{\sigma}})
\nonumber\\
&&\times (\mathbf{c}_{12}\cdot\mathbf{C}_{12}),
\end{eqnarray}
where $\mathbf{c}_{12}=\mathbf{c}_{1}-\mathbf{c}_{2}$,
$\mathbf{C}_{12}=\frac{1}{2}(\mathbf{c}_{1}+\mathbf{c}_{2})$.
Consequently,
\begin{equation}
\label{A3}
\widetilde{\Phi}_2(\mathbf{c}_{1},\mathbf{c}_{2})=
\frac{\beta_3(1-\alpha^2)}{4}c_{12}^3,
\end{equation}
\begin{eqnarray}
\label{A4}
\widetilde{\Phi}_4(\mathbf{c}_{1},\mathbf{c}_{2})
&=&\beta_3 c_{12}\left\{
\frac{(d+2)(1-\alpha^2)}{2d}c_{12}^2C_{12}^2
\right.\nonumber\\
&&
+\frac{(1-\alpha^2)(d+1+2\alpha^2)}{8(d+3)}
c_{12}^4
\nonumber\\
&&+
\frac{(2d+3-3\alpha)(1+\alpha)}{d+3}
\nonumber\\
&&\left.\times\left[
\left(\mathbf{c}_{12}\cdot
\mathbf{C}_{12}\right)^2-\frac{1}{d}c_{12}^2C_{12}^2\right]\right\}
,
\end{eqnarray}
where we have taken into account that
\begin{equation}
\label{A5}
\int
d\widehat{\boldsymbol{\sigma}}\,
\Theta(\widehat{\mathbf{c}}\cdot\widehat{\boldsymbol{\sigma}})
(\widehat{\mathbf{c}}\cdot\widehat{\boldsymbol{\sigma}})^n
=\pi^{(d-1)/2}\frac{\Gamma((n+1)/2)}{\Gamma((n+d)/2)}\equiv \beta_n,
\end{equation}
\begin{equation}
\label{A6}
\int
d\widehat{\boldsymbol{\sigma}}\,
\Theta(\widehat{\mathbf{c}}\cdot\widehat{\boldsymbol{\sigma}})
(\widehat{\mathbf{c}}\cdot\widehat{\boldsymbol{\sigma}})^n
\widehat{\boldsymbol{\sigma}}
=\beta_{n+1}\widehat{\boldsymbol{c}},
\end{equation}
\begin{equation}
\label{A7}
\int
d\widehat{\boldsymbol{\sigma}}\,
\Theta(\widehat{\mathbf{c}}\cdot\widehat{\boldsymbol{\sigma}})
(\widehat{\mathbf{c}}\cdot\widehat{\boldsymbol{\sigma}})^n
\widehat{\boldsymbol{\sigma}}
\widehat{\boldsymbol{\sigma}}
=\frac{\beta_{n}}{n+d}\left(n\widehat{\boldsymbol{c}}\widehat{\boldsymbol{c}}
+ \mathsf{I}\right).
\end{equation}
In the three-dimensional case, Eqs.\ (\ref{A1}),  (\ref{A3})
and (\ref{A4}) yield Eqs.\ (\ref{61})--(\ref{63}).

%
%

\end{document}